\documentclass[reprint,amsmath,amssymb,aps,prl,superscriptaddress]{revtex4-2}

\usepackage[utf8]{inputenc}
\usepackage{graphicx}
\usepackage{dcolumn}
\usepackage{bm}
\usepackage[space]{grffile}
\usepackage[export]{adjustbox} 
\usepackage{booktabs}
\usepackage[caption=false]{subfig}
\usepackage{natbib}
\usepackage{mathtools}

\usepackage[breaklinks,colorlinks]{hyperref}
\hypersetup{
     citecolor    = magenta,
     urlcolor = blue,
     linkcolor = blue,
}

\newcommand{\bh}{Bhattacharyya }

\begin{document}

\title{
Surveying the space of descriptions of a composite system with machine learning
}

\author{Kieran A. Murphy}
\affiliation{Dept. of Bioengineering, School of Engineering \& Applied Science, University of Pennsylvania}
\author{Yujing Zhang}
\affiliation{School of Industrial Engineering, Purdue University}
\author{Dani S. Bassett} 
\affiliation{Dept. of Bioengineering, School of Engineering \& Applied Science, University of Pennsylvania}
\affiliation{Dept. of Electrical \& Systems Engineering, School of Engineering \& Applied Science, University of Pennsylvania}
\affiliation{Dept. of Neurology, Perelman School of Medicine, University of Pennsylvania}
\affiliation{Dept. of Psychiatry, Perelman School of Medicine, University of Pennsylvania}
\affiliation{Dept. of Physics \& Astronomy, College of Arts \& Sciences, University of Pennsylvania}
\affiliation{The Santa Fe Institute}
\affiliation{Montreal Neurological Institute, McGill University}

\begin{abstract}
Multivariate information theory provides a general and principled framework for understanding how the components of a system are connected.
Existing analyses are coarse in nature---built up from characterizations of discrete subsystems---and can be computationally prohibitive.
In this work, we propose to study the continuous space of possible descriptions of a composite system as a window into its organizational structure.
A description consists of specific information conveyed about each of the components, and the space of possible descriptions is equivalent to the space of lossy compression schemes of the components.
We introduce a machine learning framework to optimize descriptions that extremize key information theoretic quantities used to characterize organization, such as total correlation and O-information.
Through case studies on spin systems, sudoku boards, and letter sequences from natural language, we identify extremal descriptions that reveal how system-wide variation emerges from individual components.
By integrating machine learning into a fine-grained information theoretic analysis of composite random variables, our framework opens a new avenues for probing the structure of real-world complex systems.
\end{abstract}

\maketitle

Multivariate information theory has emerged as a powerful lens for the understanding of complex systems, offering tools to uncover structure in the variation of multiple interacting components.
From broad explorations of the nature of complexity~\cite{tononi1994measure,nicolis2012foundations,ladyman2013complex,daniels2016quantifyingcollectivity,rosas2019oinfo,ehrlich2023complexity} to detailed investigations of specific systems such as the brain~\cite{martignon2000neural,wibral2017quantifying,varley2023partial,luppi2024pidbrain,pope2024time},
collective behavior in nature~\cite{miller2014measuring,pilkiewicz2020decoding,twomey2021redundancy,burns2022self}, 
gene regulatory networks~\cite{chan2017gene},
and toy models from condensed matter physics~\cite{sootla2017analyzing,rosas2023ogradients}, information-theoretic approaches characterize organizational structure and reveal hidden interdependencies.
These methods bridge disciplines, providing a domain-agnostic framework for quantifying how components interact to give rise to system-wide phenomena.

\begin{figure}
    \centering
    \includegraphics[width=\linewidth]{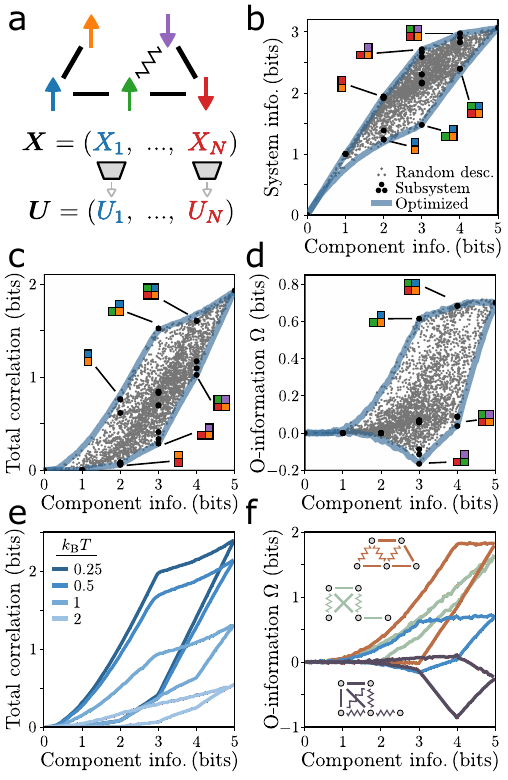}
    \caption{\textbf{Descriptions of a system.}
    \textbf{(a)} The state $\boldsymbol{X}$ of a system of five interacting spins with ferromagnetic and antiferromagnetic couplings (straight and zigzag connectors, respectively) can be \textit{described} by communicating information about each spin $X_i$.
    \textbf{(b)} The space of descriptions charted in terms of the total component information, $\sum_i I(X_i;U_i)$, and the system information, $I(\boldsymbol{X};\boldsymbol{U})$.
    Possible descriptions include the accounting of discrete subsystems---i.e., subsets of components (black circles)---as well as a continuum of compression schemes for each component, which we randomly sample (gray dots) and optimize over (blue trace, standard error visualized).
    The space of possible descriptions can also be characterized by the total correlation (\textbf{c}), O-information (\textbf{d}), or other quantities from multivariate information theory.
    \textbf{(e)} Description space in terms of total correlation for the system in panel \textbf{a} at different temperatures.
    \textbf{(f)} Description space in O-information for various five-spin systems at $k_\text{B}T=0.625$, with the blue trace the system in panel \textbf{a}.
    % For all panels, uncertainty in both the component information and the summary quantity lie within the plot lines.
    }
    \label{fig:description_space}
\end{figure}

Prior work has predominantly focused on analyses of discrete subsystems in relation to the whole system, meaning that the variation of each component in the system is considered in its entirety.
Here we propose to study the space of partial entropy allocations, which we call ``descriptions'', as a route to characterizing the interrelationships of a composite system.
A description conveys partial information about each component and can be characterized by any of the commonly used summary quantities, such as total correlation~\cite{watanabe1960information} and O-information~\cite{james2011anatomy,rosas2019oinfo}.
By formulating descriptions as a collection of communication channels, one per component (Fig.~\ref{fig:description_space}a), we can optimize descriptions with neural networks that maximize or minimize various information theoretic quantities.

We are interested in the space of possible descriptions for multiple reasons.
First, we posit that the space of descriptions is relevant to the way that humans view complex systems: due to limited processing capacity~\cite{marois2005capacity}, we focus on specific variation and ignore the rest, making the space of partial entropy allocations a natural object of study.
Consequently, the range of possible descriptions might reasonably be related to the perception of complexity~\cite{tononi1994measure}.
Second, extremal descriptions shed light on the system's  interrelationships by sifting noteworthy connections out of an abundance of variation~\cite{rosas2023ogradients}.
For example, the description with maximal total correlation~\cite{watanabe1960information} reveals the specific variation among components that is most connected.
Finally, as we will show, the space of descriptions can be navigated with machine learning, offering practicality for real-world data and the potential to scale with continued advances in machine learning.
By contrast, practicality is a significant issue for a popular framework for analyzing multivariate information content, partial information/entropy decomposition (PID/PED)~\cite{williams2010PID,ince2017ped}.
The number of PID/PED terms to evaluate grows superexponentially with system size, rendering it impractical for more than around five components~\cite{timme2014SynRedReview}.
The terms generally require exhaustive calculation, though we note a recent exception that proposes to optimize a subset of terms with machine learning~\cite{kolchinsky2024pib}.
With continuing advances in machine learning methods to compress data for revealing important variation~\cite{alemiVIB2016,koch2018natphys,dib_iclr,dib_pnas,dib_chaos}, there are new opportunities to study the structure of systems through the space of their descriptions.

Consider a system of $N$ components whose states are represented with random variables $X_i$ (Fig.~\ref{fig:description_space}a).
The full state of the system is represented with the random vector, $\boldsymbol{X}=(X_1,...,X_N)$.
Let $\boldsymbol{U}=(U_1,...,U_N)$ be a \textit{description} of the system state that conveys information about each component.
Each $U_i$ is generated from $X_i$ via a probabilistic transformation, $U_i=f_i(X_i,\epsilon_i)$, where $\epsilon_i$ is an independent noise variable that introduces stochasticity into the transformation but carries no information about any component of $\boldsymbol{X}$. As a result, $U_i$ is conditionally independent from all $X_{j \ne i}$ given $X_i$.

A description $\boldsymbol{U}$ is equivalent to a selection of entropy from each component.
The mutual information $I(X_i;U_i)$ is the amount of entropy from $X_i$ contained in $U_i$: $I(X_i;U_i)=H(X_i)-H(X_i|U_i)$, with $H(X_i)$ the Shannon entropy~\cite{cover1999elements}. 
We can view the pieces of information in the description as a new composite system derived from measurements of the original one, and then characterize the new system's information theoretic properties.

Among quantities that characterize the structure of entropy in a composite random variable, total correlation~\cite{watanabe1960information} measures the reduction in entropy when components are considered jointly versus independently, $\text{TC}(\boldsymbol{X}) = \sum_i^N H(X_i) - H(\boldsymbol{X})$.
Due to the conditional independence of each $U_i$ given $X_i$,
we have $H(\boldsymbol{U}_\mathcal{S}|\boldsymbol{X}_\mathcal{S}) = \sum_{i \in \mathcal{S}} H(U_i|X_i)$ for any subset of components $\mathcal{S}$ and can therefore evaluate the total correlation of the selected entropy in $\boldsymbol{U}$ in terms of transmitted information,
\begin{equation}
    \text{TC}(\boldsymbol{U}) = \sum_i^N I(X_i;U_i) - I(\boldsymbol{X};\boldsymbol{U}).
\end{equation}
Another quantity, O-information $\Omega$~\cite{james2011anatomy,rosas2019oinfo}, characterizes the interactions between components as dominated by redundancy ($\Omega>0$) or synergy ($\Omega<0$).
Redundant information is available from individual components, whereas synergistic information emerges only from their combinations~\cite{varley2023partial}.
For a description $\boldsymbol{U}$ of a system $\boldsymbol{X}$, the O-information is equal to
\begin{equation}
    \Omega(\boldsymbol{U}) = (N-2)I(\boldsymbol{U};\boldsymbol{X}) + \sum_i^N \left[ I(U_i;X_i) - I(\boldsymbol{U}_{/i};\boldsymbol{X}_{/i}) \right],
\end{equation}
with the notation $\boldsymbol{U}_{/i}$ indicating the set that excludes index $i$, $\{U_j : j \ne i\}$.
We seek to extremize these and other quantities that are composed entirely of entropy measurements of subsets of components.  

To find descriptions that extremize a summary quantity like total correlation, we devised a deep learning setup based on constrained communication.
Information about a component $X_i$ was transmitted to a representation $U_i$ by encoding each outcome $x_i$ to a probability distribution in latent space, $p(u_i|x_i)$.
The sum total information about all components $\sum_i I(U_i;X_i) \vcentcolon = I_\text{in}$ was set to a desired value, $\hat{I}_\text{in}$, through a loss proportional to $\vert I_\text{in}-\hat{I}_\text{in}\vert$~\cite{burgess2018understanding}.
We estimated $I(U_i;X_i)$ with a lower bound based on likelihood ratios computed from a batch of data~\cite{poole2019variational}, and then sampled latent points for the description components $u_i\sim p(u_i|x_i)$ in the remaining loss calculations.

The remaining mutual information terms include subsets with multiple components and were optimized with InfoNCE~\cite{oord2018InfoNCE}, a variant of noise contrastive estimation (NCE) common in representation learning~\cite{chen2020simclr}.
InfoNCE approximates the mutual information between two variables by contrasting positive and negative pairs of their outcomes.
This process employs two neural networks: one to encode the first variable and another to encode the second, mapping them into a shared representation space.
The InfoNCE loss for a batch of $B$ samples, indexed by superscripts $\alpha$ and $\beta$, is given by:
\begin{equation}
    \mathcal{L}_\text{NCE}(U;X) = -\frac{1}{B} \sum_{\alpha=1}^B \log \frac{\exp(s(u^\alpha,x^\alpha))}{\sum_{\beta=1}^B \exp(s(u^\alpha,x^\beta))},
\end{equation}
where $s(u^\alpha,x^\beta)$ measures the similarity between the representations of $u^\alpha$ and $x^\beta$ in the shared space, taken to be the squared Euclidean distance in this work.
Positive pairs $(u^\alpha,x^\alpha)$ are contrasted against negative pairs $(u^\alpha,x^\beta)$ sampled within the batch.

To summarize, the training loss combines \textbf{(i)} a sum total information constraint on the pieces of the description with \textbf{(ii)} any remaining terms in the summary quantity that we wish to extremize.
For example, to minimize total correlation would require minimizing $\mathcal{L}=\gamma \vert I_\text{in}-\hat{I}_\text{in}\vert + \mathcal{L}_\text{NCE}(\boldsymbol{U};\boldsymbol{X})$.
We note that this formulation closely resembles a distributed information bottleneck (IB) scenario with the joint variable $\boldsymbol{X}$ serving as the relevance variable~\cite{ince2017ped,dib_pnas}.

To maximize total correlation requires minimizing $I(\boldsymbol{U};\boldsymbol{X})$, and InfoNCE, as a lower bound, cannot be directly used.
For any such minimization term, we employed an adversarial setup where auxiliary networks were trained to maximize mutual information via InfoNCE, and then the loss was negated and applied to the description encodings.

For evaluation, mutual information terms were estimated via Monte Carlo sampling using likelihood ratios. Standard error is reported and lies within plot markers for all results presented in this work.

We examined the space of descriptions for three systems.
A spin system with $N=5$ sites has ferromagnetic and antiferromagnetic couplings (Fig.~\ref{fig:description_space}a).
The probability distribution over states $\boldsymbol{x}\!=\!(x_1, ..., x_N)$ is given by
\begin{equation}
    p(\boldsymbol{x}) = \frac{1}{Z} \exp (-\mathcal{H}(\boldsymbol{x})/k_\text{B}T).
\end{equation}
$Z$ is a normalization constant called the partition function and $\mathcal{H}(\boldsymbol{x})=-\sum_{\langle ij \rangle} J_{ij} x_i x_j$ is the energy of the state $\boldsymbol{x}$.
$J_{ij}=1$, $-1$ for the ferromagnetic and antiferromagnetic couplings, and $0$ otherwise; we set $k_\text{B}T=0.625$.

In Fig.~\ref{fig:description_space}b-d, we surveyed the space of descriptions for the 5-spin Ising system in several ways.
First, descriptions were randomly sampled by creating a binary symmetric channel~\cite{cover1999elements} for each spin with random noise (gray dots).
Second, we formed descriptions corresponding to complete information about each possible discrete subsystem (black circles).
Third, we probed the boundaries by extremizing total correlation or O-information (light blue curve). 
The descriptions that extremize total correlation also extremize system information (Fig.~\ref{fig:description_space}b).
The method successfully found descriptions that closely trace the bounds of the randomly sampled descriptions, which can be densely sampled for this small system.

We interpret the space of descriptions \textit{globally} through its overall shape when charted in terms of different quantities, and \textit{locally} through specific extremal descriptions.
From a global perspective, the space of descriptions can tell of the nature of interactions between components across different levels of information.
The spin system in Fig.~\ref{fig:description_space}a has three-bit descriptions that are either redundant or synergistic, even though the full state description is redundant (Fig.~\ref{fig:description_space}d).
Intuitively, the range of descriptions narrows and drops in total correlation as the temperature of the system grows (Fig.~\ref{fig:description_space}e).
With different connections between the spins, description space boundaries can vary dramatically and highlight qualitatively distinct multivariate relationships (Fig.~\ref{fig:description_space}f).

From a local perspective, the extremal descriptions reveal subsystems of interest.
Among three-bit descriptions, the maximal total correlation for the five spin system in Fig.~\ref{fig:description_space}a links the ferromagnetic chain (orange, blue, green) (Fig.~\ref{fig:description_space}c).
The minimal total correlation connects the most distant spins (orange and red), then incorporates a second spin (purple) from the frustrated triangle---i.e., three spins connected by couplings for which there must be at least one inconsistent edge.
In terms of O-information, maximal redundancy lies in the ferromagnetic chain while maximal synergy resides in the frustrated triangle (Fig.~\ref{fig:description_space}d).
For the case of binary variables, broadening the space of descriptions from discrete subsystems to include partial information merely fills in the space between subsystem descriptions and reveals nothing new, though it facilitates optimization.

\begin{figure}
    \centering
    \includegraphics[width=\linewidth]{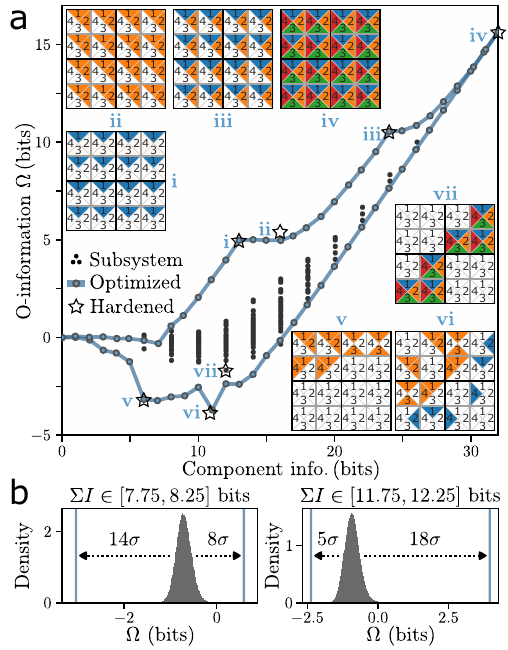}
    \caption{\textbf{The space of descriptions of a 4x4 sudoku board.}
    \textbf{(a)} Discrete subsets of squares (black circles) and machine learning-optimized boundaries (blue circles), in terms of O-information.
    Optimized soft compression schemes are converted to hard compression schemes (black stars) and visualized according to the corresponding Roman numerals.
    The hard compression scheme for each square in a board is displayed by coloring numbers according to groupings.
    For example, if one number in a square is blue and the rest are white, the blue number is distinguishable from the remaining three, and the three are indistinguishable from each other.
    \textbf{(b)} We randomly sampled $10^6$ hard descriptions within the information range at the top of each plot. 
    The optimized descriptions have O-information values (blue vertical lines) far from the distribution of randomly sampled schemes (grey).
    }
    \label{fig:sudoku}
\end{figure}

The second system is a 4x4 sudoku board (Fig.~\ref{fig:sudoku}), where every square contains a digit from one to four and no digit can be repeated in a row, column, or quadrant. 
Sudoku is representative of constraint satisfaction problems, which are rich in structure and central to a wide variety of practical and theoretical challenges, from scheduling to combinatorial optimization~\cite{gomes2000heavy,ercsey2012chaos,varga2016order}. 
We took the probability of states $p(\boldsymbol{x})$ to be uniform across valid boards and zero otherwise. 
The dense constraints severely restrict the number of valid boards from $4^{16}$ ($\approx10^9$) to 288, suggesting an intricate organizational structure linking the states of the squares.
For this and remaining analyses, we found it necessary to run a different optimization per value of $\hat{I}_\text{in}$, in contrast to the spin system where a single optimization spanned the full range of description information $\hat{I}_\text{in}$.
For each optimization, the coefficient of the transmitted information loss, $\gamma$, was constant for the first half of training, and then increased exponentially over the second half. 

We focused on O-information to highlight modes of information sharing (Fig.~\ref{fig:sudoku}a); analysis using total correlation, more revealing of component (in)dependence, is in the Supp.
All descriptions become highly redundant after around 16 bits, reflective of the dense constraints between squares.
The discrete subsystems of a board state have minimal O-information (maximal synergy) at six squares and include the distantly connected triplets of squares shown in scheme \textbf{vii}.

In contrast to the spin system, the machine learning-optimized boundaries for the sudoku board reach beyond the descriptions of discrete subsystems.
The optimized compression schemes are soft, meaning that the information conveyed about each square resides in the layout of distributions $p(u_i|x_i)$ in latent space and communicates partial distinguishability between possible outcomes.
For ease of interpretation, we converted each square's compression scheme to a hard clustering of outcomes.
After optimization, gradient descent was performed directly on the compression scheme to drive the statistical distinguishability between probabilistic representations to perfect distinguishability/indistinguishability, as measured by the \bh coefficient~\cite{kailath1967Bdistance,dib_iclr,nmi2024}. 

The hardened compression schemes show which digits for each square were clustered together (having the same color in the insets to Fig.~\ref{fig:sudoku}a), and are intuitive for the maximal O-information (redundant) descriptions.
The same partial information is communicated about every square, from 0.8 bits per square to distinguish one number from the other three (descr. \textbf{i}) to distinguishing all four digits (descr. \textbf{iv}).
The descriptions with minimal O-information (synergistic) (\textbf{v}-\textbf{vii}) are less comprehensible.
Description \textbf{v} communicates one bit three different ways, spread out across a row and a quadrant.
The most synergistic description found for any level of communicated information, \textbf{vi}, is an intricate pattern of partial information, with a symmetry across the diagonal that was also present in the subsystem with minimal O-information (descr. \textbf{vii}).

The space of hard compression schemes of a discrete random variable with $m$ outcomes is equivalent to the partitions of a set of $m$ elements.
There are 15 possible hard compression schemes for a variable with four outcomes, and 16 squares, making $15^{16}\approx10^{18}$ different hard descriptions of a 4x4 sudoku board.
Without accounting for symmetries, a manual search through these descriptions is impractical; by contrast, the machine learning approach identified extremal descriptions on the order of a few minutes per optimization.
To ground the performance of the extremized descriptions, we employed rejection sampling to obtain $10^6$ random hard descriptions for two ranges of component information (Fig.~\ref{fig:sudoku}b).
The extremized descriptions are several standard deviations away from the mean O-information of the sampled descriptions.
Importantly, while there is no guarantee of global optimality, extremized descriptions bound the optimum, can be improved through hyperparameter tuning or repeated runs, and nevertheless highlight notable entropy allocations.

\begin{figure}
    \centering
    \includegraphics[width=\linewidth]{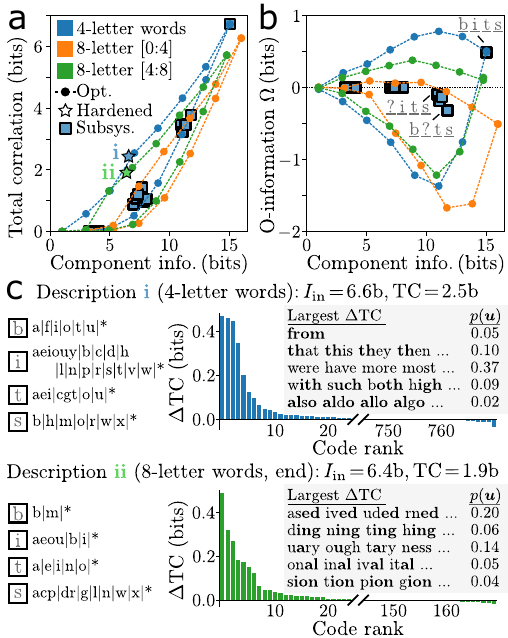}
    \caption{\textbf{Statistical structure in 4-letter sequences.}
    The space of descriptions for 4-grams taken from 4- and 8-letter words, plotted in terms of \textbf{(a)} total correlation and \textbf{(b)} O-information.
    \textbf{(c)} The hardened descriptions for the maximal total correlation points marked as stars in panel \textbf{(a)}, where groupings of letters are separated by vertical bars, and the group with all remaining letters of the alphabet is represented by an asterisk (*).
    The top contributions $\Delta\text{TC}$ to total correlation are shown at right, with the most probable n-grams inside each grouping shown.  
    Letters are bolded to highlight recognizable letter patterns central to each grouping. 
    }
    \label{fig:words}
\end{figure}

Finally, we analyzed the statistics of 4-grams in the English language based on data from Wikipedia.
Letters are combined with numerous soft constraints that facilitate learning and error correction, and have long been used as an object of study in information theory~\cite{shannon1948mathematical} and statistical mechanics~\cite{stephens2010letters}.
We surveyed the space of descriptions of 4-grams that are themselves 4-letter words and the first or second half of 8-letter words.
From a global perspective, total correlation varied less across the different 4-grams than did O-information (Fig.~\ref{fig:words}a,b).
The space of descriptions is almost entirely synergistic for the first four letters of 8-letter words, whereas 4-letter words contain variation that is more redundant.
A typical O-information assessment, without the notion of a description and corresponding to full information about each letter, would also be negative for the former and positive for the latter (rightmost points).
The most negative O-information occurs around 10–12 bits of component information---well beyond the full-information value.
Discrete subsystems, shown for the 4-letter words in Fig.~\ref{fig:words}a,b (squares), are far too coarse to capture the shape of the space of descriptions.

We hardened optimized descriptions (stars in Fig.~\ref{fig:words}a), and focus on maximal total correlation here due to its relative interpretability (other hardened descriptions in the Supp).
After hardening, each 4-gram cluster corresponds to a particular code $\boldsymbol{u}$ and contributes $\Delta\text{TC}(\boldsymbol{u}) \coloneqq p(\boldsymbol{u}) \text{tc}(\boldsymbol{u})$, where $\text{tc}(\boldsymbol{u})$ is the local (pointwise) total correlation~\cite{scagliarini2022local} and $\text{TC}(\boldsymbol{U})=\sum_u \Delta\text{TC}(\boldsymbol{u})$ with $u \in \mathcal{U}$ the set of possible codes.
Evidently, to maximize total correlation, it is important to group 4-letter words starting with \texttt{th}, and the second half of 8-letter words that end in \texttt{ed} and \texttt{ing}.
These schemes provide a starting point for a deeper linguistic analysis, serving as a sieve on the space of groupings of letters.

In this work, we introduced a machine learning framework to study the space of partial descriptions of composite systems---a space too vast to adequately explore, even by random sampling, for all but the simplest cases. 
Crucially, each learned mapping from $\boldsymbol{X}$ to $\boldsymbol{U}$ defines a valid description of the system, so the method does not depend on convergence in the conventional sense: even suboptimal solutions yield interpretable and potentially informative points in the space of descriptions.
Additionally, though our focus was on discrete variables, the approach extends naturally to continuous ones.
Finally, the framework is flexible enough to extremize a broad class of quantities, such as the binding entropy~\cite{james2011anatomy}, S-information~\cite{james2011anatomy,rosas2019oinfo,varley2023partial} and $\Delta I$~\cite{nirenberg2001retinal}, Tononi-Sporns-Edelman complexity~\cite{tononi1994measure}, and specific atoms of PED~\cite{ince2017ped}, with each offering a different view of how entropy is distributed across components.
Altogether, the framework open new avenues for exploring the structural underpinnings of complex systems, offering both theoretical insights and practical tools for uncovering a scaffold of interconnected variation across components.

\clearpage

\onecolumngrid

\setcounter{section}{0}
\setcounter{page}{1}
\setcounter{figure}{0}
\setcounter{table}{0}

\renewcommand{\thepage}{S\arabic{page}}
\renewcommand{\thesection}{S\arabic{section}}
\renewcommand{\thetable}{S\arabic{table}}
\renewcommand{\thefigure}{S\arabic{figure}}
\renewcommand{\figurename}{Supplemental Material, Figure}

\hrule
\vspace{3mm}
{\Large Supplemental Material }
\vspace{3mm}

\noindent The code base has been released on Github at \href{https://github.com/murphyka/description\_space}{https://github.com/murphyka/description\_space}.
The analyses of the three systems from the main text can be repeated with the training script in the linked repository.
Experiments were implemented in TensorFlow and run on a single computer with a 12 GB GeForce RTX 3060 GPU. 

\noindent \textbf{Data.} We scraped the 288 valid 4x4 sudoku boards from \href{https://sudokuprimer.com/4x4puzzles.php}{https://sudokuprimer.com/4x4puzzles.php} and include the valid boards in this project's github repo.
For the 4-gram analysis, we downloaded word frequency data from \href{https://github.com/IlyaSemenov/wikipedia-word-frequency}{https://github.com/IlyaSemenov/wikipedia-word-frequency}; there, the file \texttt{results/enwiki-2023-04-13.txt} has frequency counts from English Wikipedia dated April 13, 2023.
We truncated the length 4 and length 8 frequency tables after 10,000 entries, and then further discarded any entries that included a symbol outside of the 26 letters (a-z).
When compiling statistics for the beginning and ending 4-grams inside length 8 words, there can be duplicates (e.g. \texttt{info} as a part of \texttt{informal} and \texttt{informed}); we combined the frequency counts for any such duplicates.

\noindent \textbf{Specifying the quantity to extremize.}
The proposed method can optimize descriptions that extremize any summary quantity composed of mutual information terms of the form $I(\{U_i\}_{i \in \mathcal{A}};\{X_i\}_{i \in \mathcal{A}})$, where $\mathcal{A}$ is a set of indices of components in that term.
In the project's codebase, the information theoretic quantity to extremize is specified with a list of lists---with each inner list representing one such mutual information term---and a corresponding list of weights.
For example, consider a system with three components.
The total correlation of a description $\boldsymbol{U}=(U_1, U_2, U_3)$ is
\begin{equation}
    \text{TC}(\boldsymbol{U})=I(U_1;X_1)+I(U_2;X_2)+I(U_3;X_3)-I(U_1,U_2,U_3;X_1,X_2,X_3).
\end{equation}
The mutual information terms can be specified as the list of lists \texttt{[[1], [2], [3], [1, 2, 3]]}, with corresponding weights \texttt{[1, 1, 1, -1]} if total correlation is to be minimized.
To maximize total correlation, one simply negates all weights.
To maximize O-information over the same three component system, one would require the terms \texttt{[[0], [1], [2], [1, 2], [0, 2], [0, 1], [0, 1, 2]]} and weights \texttt{[-1, -1, -1,  1,  1,  1, -1]}.

\noindent The codebase is written to expect the first $N$ terms to be the individual components, as these have a special role in the training loss for two reasons.
First, the sum total information from all components is driven to a specified value $\hat{I}_\text{in}$, rather than extremized as is done for the remaining terms.
Second, the component-wise information terms $I(U_i;X_i)$ are estimated with the lower bound in Section 2.5 of \citet{poole2019variational}, while the remaining mutual information terms are estimated through InfoNCE~\cite{oord2018InfoNCE}.

\noindent \textbf{The training process.}
For the spin system, a single optimization was run for all transmitted information values $\hat{I}_\text{in}$ for each extremization (minimization or maximization) of a given quantity.
The value was increased linearly from zero bits to five bits over the course of training.
For sudoku and the $n$-gram statistics, we found it necessary to run separate optimizations for each value of $\hat{I}_\text{in}$, and to increase the coefficient $\gamma$ on the sum total transmitted information in stages over the course of training.
During each optimization, $\gamma$ was held fixed at a low value $\gamma_0$ for the first half of training.
Then it was increased exponentially to its final value $\gamma_1$ over the second half of training.

\noindent For evaluation, we sampled data points $x\sim p(x)$ and then embeddings $u\sim p(u|x)$ to compute the expectation \begin{equation}
    I(X;U)=\mathbb{E}_{x,u\sim p(x,u)} [\log \frac{p(u|x)}{p(u)}],
\end{equation}
with $p(u)=\sum_i^M p(x_i) p(u|x_i)$ aggregated over the entire dataset.
For all analyses, we sampled $2\times10^5$ points and used the standard error of the estimate as its uncertainty.  
Error propagation then gave the uncertainties on the summary quantities and total component information.

\noindent For the spin systems and sudoku, we repeated each run five times and used the best performer, which was simply the point (or scan) that yielded the maximal/minimal summary quantity.
For n-grams, we trained once, albeit after some hyperparameter tuning.

\noindent \textbf{Training hyperparameters and architecture details.} For each of the three systems, we list implementation specifics in Tables S1-3.
The ``Encoder MLP architecture'' was used for every MLP involved in the InfoNCE terms (two MLPs per term).

\begin{table}
\centering
\begin{tabular}{||c c||} 
 \hline
 Parameter & Value \\ 
 \hline\hline
 Bottleneck embedding space dimension & 2 \\
 Encoder MLP architecture & [256 \texttt{leaky\_ReLU}] \\
 InfoNCE similarity metric $s(u,v)$ & Euclidean squared\\
 InfoNCE space dimensionality & 32\\
 InfoNCE temperature & 1\\
 \hline 
 Batch size & 256 \\
 Optimizer, latent encodings & SGD \\
 Learning rate, latent encodings & $1 \times 10^{-2}$ \\
 Optimizer, InfoNCE & Adam \\
 Learning rate, InfoNCE & $3 \times 10^{-4}$ \\
 
 Transmitted information coefficient $\gamma$ & 1 \\
 
 Training steps & $5 \times 10^4$\\
 Further InfoNCE optimization steps & $2 \times 10^4$\\
 
 \hline
\end{tabular}
\caption{Training parameters for the five spin system.}
\label{tab:hparams_spins}
\end{table}

\begin{table}
\centering
\begin{tabular}{||c c||} 
 \hline
 Parameter & Value \\ 
 \hline\hline
 Bottleneck embedding space dimension & 2 \\
 Encoder MLP architecture & [512 \texttt{leaky\_ReLU}, 512 \texttt{leaky\_ReLU}] \\
 InfoNCE similarity metric $s(u,v)$ & Euclidean squared\\
 InfoNCE space dimensionality & 32\\
 InfoNCE temperature & 1\\
 \hline 
 Batch size & 576 \\
 Optimizer, latent encodings & SGD \\
 Learning rate, latent encodings & $1 \times 10^{-2}$ \\
 Optimizer, InfoNCE & Adam \\
 Learning rate, InfoNCE & $1 \times 10^{-4}$ \\
 
 Transmitted information coefficient $\gamma_0$ & 1 \\
 Transmitted information coefficient $\gamma_1$ & 10 \\
 
 Training steps & $2 \times 10^4$\\
 
 \hline
\end{tabular}
\caption{Training parameters for 4$\times$4 sudoku.}
\label{tab:hparams_sudoku}
\end{table}

\begin{table}
\centering
\begin{tabular}{||c c||} 
 \hline
 Parameter & Value \\ 
 \hline\hline
 Bottleneck embedding space dimension & 8 \\
 Encoder MLP architecture & [256 \texttt{leaky\_ReLU}, 256 \texttt{leaky\_ReLU}, 256 \texttt{leaky\_ReLU}] \\
 InfoNCE similarity metric $s(u,v)$ & Euclidean squared\\
 InfoNCE space dimensionality & 32\\
 InfoNCE temperature & 1\\
 \hline 
 Batch size & 1024 \\
 Optimizer, latent encodings & SGD \\
 Learning rate, latent encodings & $1 \times 10^{-2}$ \\
 Optimizer, InfoNCE & Adam \\
 Learning rate, InfoNCE & $3 \times 10^{-4}$ \\
 
 Transmitted information coefficient $\gamma_0$ & 2 \\
 Transmitted information coefficient $\gamma_1$ & 10 \\
 
 Training steps & $2 \times 10^5$\\
 
 \hline
\end{tabular}
\caption{Training parameters for $n$-grams.}
\label{tab:hparams_words}
\end{table}

\begin{figure}
    \centering
    \includegraphics[width=0.45\linewidth]{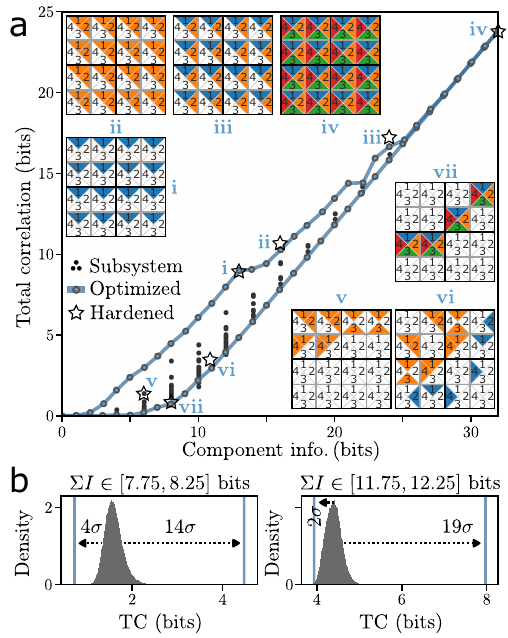}
    \caption{\textbf{The space of descriptions of a 4x4 sudoku board in terms of total correlation.}
    Discrete subsets of squares (black circles) and machine learning-optimized boundaries (blue curves), in terms of total correlation.
    Optimized (soft) compression schemes are converted to hard compression schemes (black stars) and visualized according to the corresponding Roman numerals.
    The hard compression scheme for each square in a board is displayed by coloring numbers according to groupings.
    For example, if one number in a square is blue and the rest are white, the blue number is distinguishable from the remaining three, and the three are indistinguishable from each other.
    }
    \label{fig:sudokutc}
\end{figure}

\noindent \textbf{Sudoku description space in terms of total correlation.}
Fig.~\ref{fig:sudokutc} is the counterpart to Fig.~\ref{fig:sudoku} in the main text, where the description space for sudoku is visualized and extremized in terms of total correlation instead of O-information.

\noindent The most redundant descriptions from before are also the ones with largest total correlation.
However, the most synergistic descriptions found before (schemes \textbf{v} and \textbf{vi}) lie internal to the total correlation boundaries.
The subsystem with minimal total correlation for eight bits of information (four squares), scheme \textbf{vii}, shows the most independent squares on the board.

\begin{figure}
    \centering
    \includegraphics[width=\linewidth]{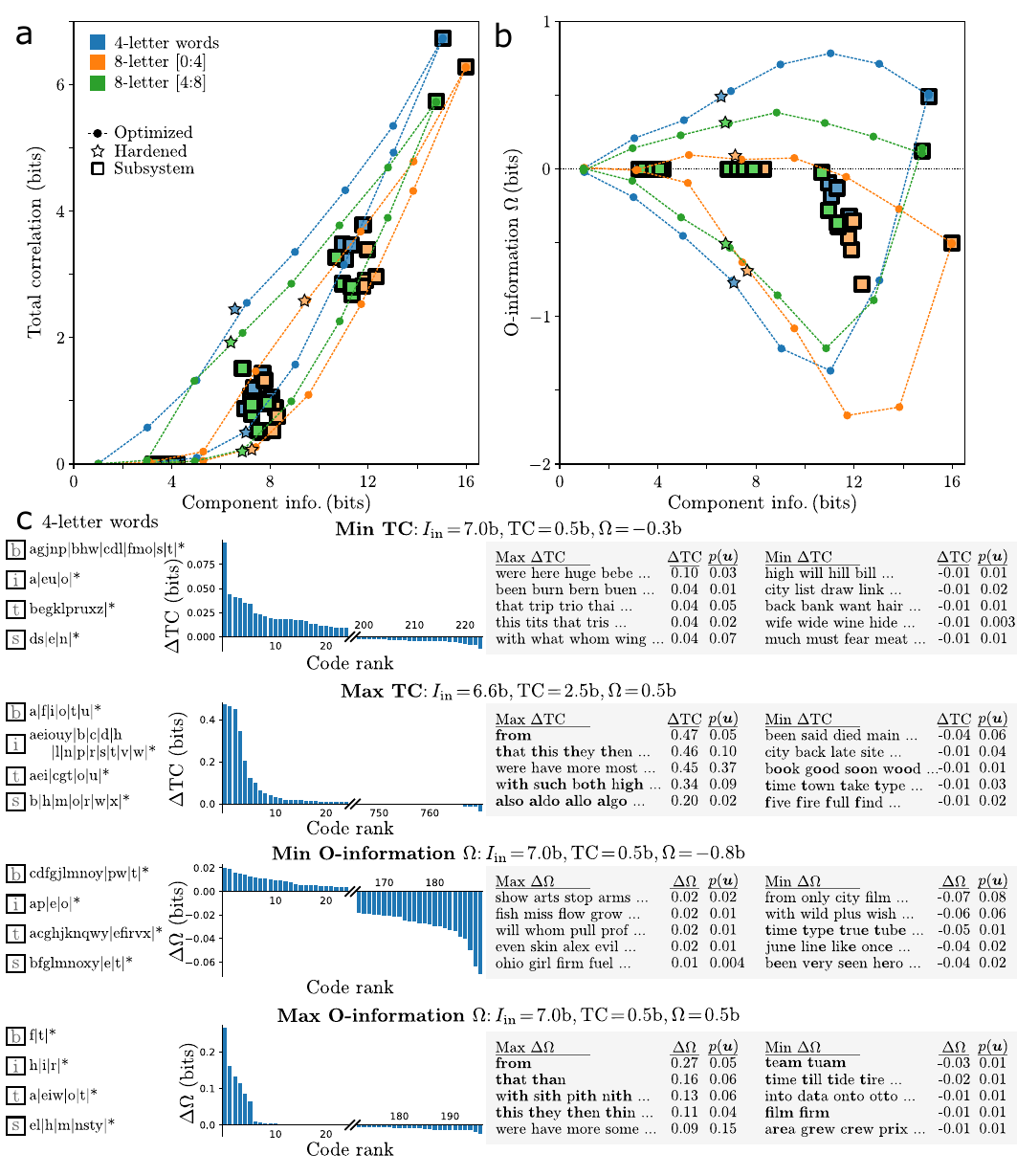}
    \caption{\textbf{Structure of 4-grams statistics, continued.}
    We reproduce the total correlation \textbf{(a)} and O-information \textbf{(b)} description spaces from Fig.~\ref{fig:words} in the main text, now with the discrete subsystems for all three datasets (squares), and with hardened descriptions for all extremized quantities at around seven bits of total information (stars).
    \textbf{(c)} For the 4-letter words, we hardened the descriptions that minimize and maximize total correlation and O-information, and show the top contributing codes.
    }
    \label{fig:ngrams_supp1}
\end{figure}

\begin{figure}
    \centering
    \includegraphics[width=\linewidth]{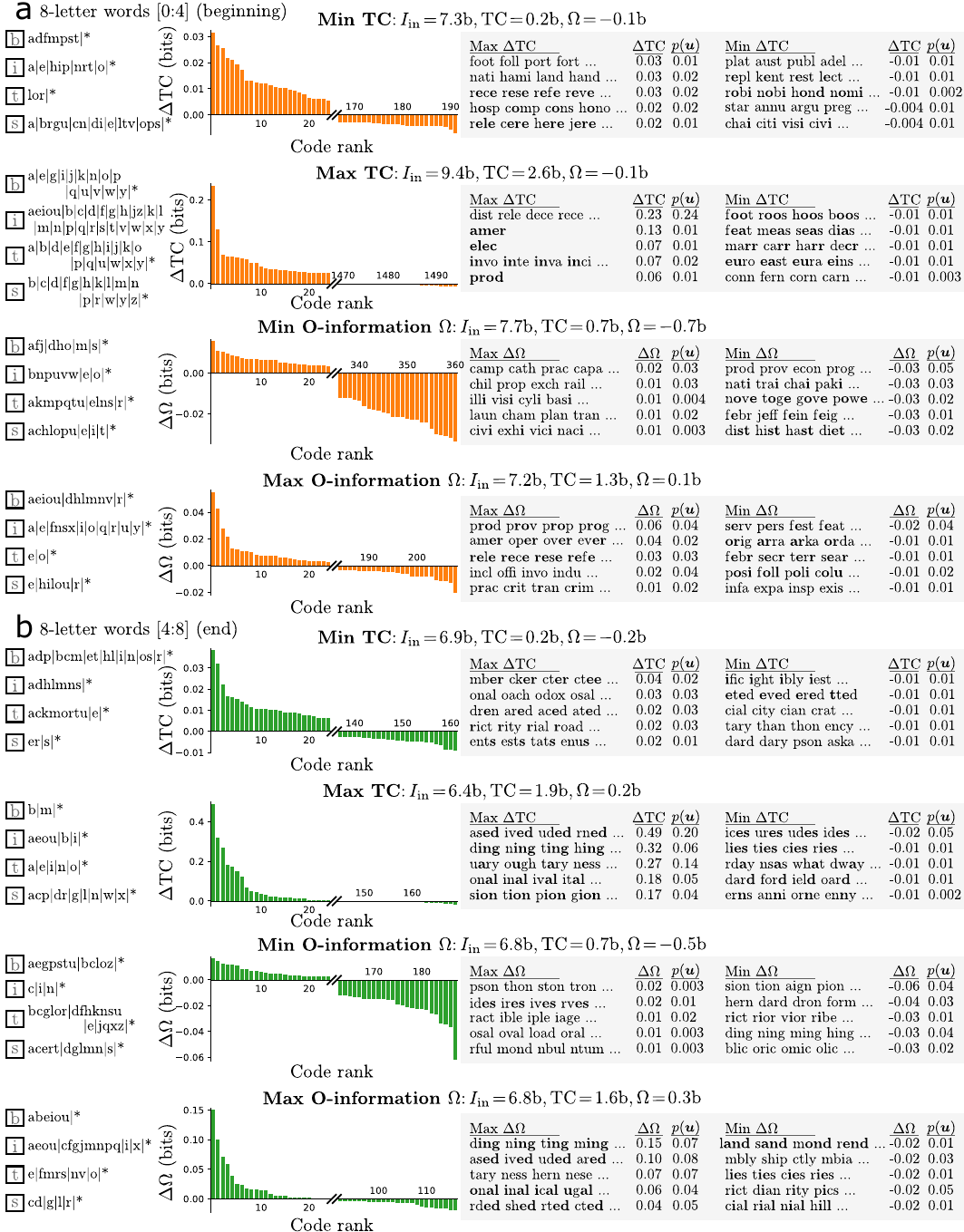}
    \caption{\textbf{Structure of 4-grams statistics, continued.}
    For the first half \textbf{(a)} and second half \textbf{(b)} of 8-letter words, we hardened the descriptions that minimize and maximize total correlation and O-information, and show the top contributing codes.
    }
    \label{fig:ngrams_supp2}
\end{figure}

\noindent \textbf{Contribution of specific codes to summary quantities (Fig.~\ref{fig:words}, main text).}
By expressing total correlation $\text{TC}(\boldsymbol{U})$ as an expectation over codes $\boldsymbol{u}$, we can compute the contribution per code for additional insight about the variation that a description encapsulates~\cite{watanabe1960information,scagliarini2022local}.
To be specific, 
\begin{equation}
    \text{TC}(\boldsymbol{U}) = \mathbb{E}_{\boldsymbol{u}\sim p(\boldsymbol{u})}[\log \frac{p(\boldsymbol{u})}{\Pi_i^N p(u_i)}] = \mathbb{E}_{\boldsymbol{u}\sim p(\boldsymbol{u})} [\text{tc}(\boldsymbol{u})],
\end{equation}
framing total correlation as a comparison between the probability of outcome $\boldsymbol{u}$ when accounting for all components jointly, $p(\boldsymbol{u})$, versus independently, $\Pi_i^N p(u_i)$.
We then evaluate the contribution of each outcome to total correlation as $\Delta \text{TC}(\boldsymbol{u}) = p(\boldsymbol{u}) \cdot \text{tc}(\boldsymbol{u})$.

\noindent O-information permits a similar framing~\cite{scagliarini2022local}, now as a comparison between the joint and product-of-marginals probabilities, and the joint and marginalize-one-out probabilities:
\begin{equation}
    \Omega(\boldsymbol{U}) = \mathbb{E}_{\boldsymbol{u}\sim p(\boldsymbol{u})}[2\log \frac{p(\boldsymbol{u})}{\Pi_i^N p(u_i)}-\sum_i^N \log \frac{p(\boldsymbol{u})}{p(\boldsymbol{u}_{/i})p(u_i)}] = \mathbb{E}_{\boldsymbol{u}\sim p(\boldsymbol{u})} [\omega(\boldsymbol{u})].
\end{equation}
We show the sorted contributions to total correlation ($\Delta$TC) and O-information ($\Delta\Omega$), and the groupings of 4-grams that contribute most on either end of the spectrum, in Figs.~\ref{fig:ngrams_supp1}\&\ref{fig:ngrams_supp2}.

\section{Citation Diversity Statement}
Science is a human endeavour and consequently vulnerable to many forms of bias; the responsible scientist identifies and mitigates such bias wherever possible.
Meta-analyses of research in multiple fields have measured significant bias in how research works are cited, to the detriment of scholars in minority groups~\citep{maliniak2013gender,caplar2017quantitative,chakravartty2018communicationsowhite,dion2018gendered,dworkin2020extent,teich2022citation}.
We use this space to amplify studies, perspectives, and tools that we found influential during the execution of this research~\citep{zurn2020citation,dworkin2020citing,zhou2020gender,budrikis2020growing}.
We sought to proactively consider choosing references that reflect the diversity of the field in thought, form of contribution, gender, race, ethnicity, and other factors. 
The gender balance of papers cited within this work was quantified using a combination of automated \href{https://gender-api.com}{\color{blue}gender-api.com} estimation and manual gender determination from authors’ publicly available pronouns.
By this measure (and excluding self-citations to the first and last authors of our current paper), the references of the main text contain 5\% woman(first)/woman(last), 10\% man/woman, 18\% woman/man, and 68\% man/man. 
This method is limited in that a) names, pronouns, and social media profiles used to construct the databases may not, in every case, be indicative of gender identity and b) it cannot account for intersex, non-binary, or transgender people.
We look forward to future work that could help us to better understand how to support equitable practices in science.


\begin{thebibliography}{54}%
\makeatletter
\providecommand \@ifxundefined [1]{%
 \@ifx{#1\undefined}
}%
\providecommand \@ifnum [1]{%
 \ifnum #1\expandafter \@firstoftwo
 \else \expandafter \@secondoftwo
 \fi
}%
\providecommand \@ifx [1]{%
 \ifx #1\expandafter \@firstoftwo
 \else \expandafter \@secondoftwo
 \fi
}%
\providecommand \natexlab [1]{#1}%
\providecommand \enquote  [1]{``#1''}%
\providecommand \bibnamefont  [1]{#1}%
\providecommand \bibfnamefont [1]{#1}%
\providecommand \citenamefont [1]{#1}%
\providecommand \href@noop [0]{\@secondoftwo}%
\providecommand \href [0]{\begingroup \@sanitize@url \@href}%
\providecommand \@href[1]{\@@startlink{#1}\@@href}%
\providecommand \@@href[1]{\endgroup#1\@@endlink}%
\providecommand \@sanitize@url [0]{\catcode `\\12\catcode `\$12\catcode
  `\&12\catcode `\#12\catcode `\^12\catcode `\_12\catcode `\%12\relax}%
\providecommand \@@startlink[1]{}%
\providecommand \@@endlink[0]{}%
\providecommand \url  [0]{\begingroup\@sanitize@url \@url }%
\providecommand \@url [1]{\endgroup\@href {#1}{\urlprefix }}%
\providecommand \urlprefix  [0]{URL }%
\providecommand \Eprint [0]{\href }%
\providecommand \doibase [0]{https://doi.org/}%
\providecommand \selectlanguage [0]{\@gobble}%
\providecommand \bibinfo  [0]{\@secondoftwo}%
\providecommand \bibfield  [0]{\@secondoftwo}%
\providecommand \translation [1]{[#1]}%
\providecommand \BibitemOpen [0]{}%
\providecommand \bibitemStop [0]{}%
\providecommand \bibitemNoStop [0]{.\EOS\space}%
\providecommand \EOS [0]{\spacefactor3000\relax}%
\providecommand \BibitemShut  [1]{\csname bibitem#1\endcsname}%
\let\auto@bib@innerbib\@empty
%</preamble>
\bibitem [{\citenamefont {Tononi}\ \emph {et~al.}(1994)\citenamefont {Tononi},
  \citenamefont {Sporns},\ and\ \citenamefont {Edelman}}]{tononi1994measure}%
  \BibitemOpen
  \bibfield  {author} {\bibinfo {author} {\bibfnamefont {G.}~\bibnamefont
  {Tononi}}, \bibinfo {author} {\bibfnamefont {O.}~\bibnamefont {Sporns}},\
  and\ \bibinfo {author} {\bibfnamefont {G.~M.}\ \bibnamefont {Edelman}},\
  }\bibfield  {title} {\bibinfo {title} {A measure for brain complexity:
  relating functional segregation and integration in the nervous system.},\
  }\href@noop {} {\bibfield  {journal} {\bibinfo  {journal} {Proceedings of the
  National Academy of Sciences}\ }\textbf {\bibinfo {volume} {91}},\ \bibinfo
  {pages} {5033} (\bibinfo {year} {1994})}\BibitemShut {NoStop}%
\bibitem [{\citenamefont {Nicolis}\ and\ \citenamefont
  {Nicolis}(2012)}]{nicolis2012foundations}%
  \BibitemOpen
  \bibfield  {author} {\bibinfo {author} {\bibfnamefont {G.}~\bibnamefont
  {Nicolis}}\ and\ \bibinfo {author} {\bibfnamefont {C.}~\bibnamefont
  {Nicolis}},\ }\href@noop {} {\emph {\bibinfo {title} {Foundations of complex
  systems: emergence, information and predicition}}}\ (\bibinfo  {publisher}
  {World Scientific},\ \bibinfo {year} {2012})\BibitemShut {NoStop}%
\bibitem [{\citenamefont {Ladyman}\ \emph {et~al.}(2013)\citenamefont
  {Ladyman}, \citenamefont {Lambert},\ and\ \citenamefont
  {Wiesner}}]{ladyman2013complex}%
  \BibitemOpen
  \bibfield  {author} {\bibinfo {author} {\bibfnamefont {J.}~\bibnamefont
  {Ladyman}}, \bibinfo {author} {\bibfnamefont {J.}~\bibnamefont {Lambert}},\
  and\ \bibinfo {author} {\bibfnamefont {K.}~\bibnamefont {Wiesner}},\
  }\bibfield  {title} {\bibinfo {title} {What is a complex system?},\
  }\href@noop {} {\bibfield  {journal} {\bibinfo  {journal} {European Journal
  for Philosophy of Science}\ }\textbf {\bibinfo {volume} {3}},\ \bibinfo
  {pages} {33} (\bibinfo {year} {2013})}\BibitemShut {NoStop}%
\bibitem [{\citenamefont {Daniels}\ \emph {et~al.}(2016)\citenamefont
  {Daniels}, \citenamefont {Ellison}, \citenamefont {Krakauer},\ and\
  \citenamefont {Flack}}]{daniels2016quantifyingcollectivity}%
  \BibitemOpen
  \bibfield  {author} {\bibinfo {author} {\bibfnamefont {B.~C.}\ \bibnamefont
  {Daniels}}, \bibinfo {author} {\bibfnamefont {C.~J.}\ \bibnamefont
  {Ellison}}, \bibinfo {author} {\bibfnamefont {D.~C.}\ \bibnamefont
  {Krakauer}},\ and\ \bibinfo {author} {\bibfnamefont {J.~C.}\ \bibnamefont
  {Flack}},\ }\bibfield  {title} {\bibinfo {title} {Quantifying collectivity},\
  }\href@noop {} {\bibfield  {journal} {\bibinfo  {journal} {Current opinion in
  neurobiology}\ }\textbf {\bibinfo {volume} {37}},\ \bibinfo {pages} {106}
  (\bibinfo {year} {2016})}\BibitemShut {NoStop}%
\bibitem [{\citenamefont {Rosas}\ \emph {et~al.}(2019)\citenamefont {Rosas},
  \citenamefont {Mediano}, \citenamefont {Gastpar},\ and\ \citenamefont
  {Jensen}}]{rosas2019oinfo}%
  \BibitemOpen
  \bibfield  {author} {\bibinfo {author} {\bibfnamefont {F.~E.}\ \bibnamefont
  {Rosas}}, \bibinfo {author} {\bibfnamefont {P.~A.~M.}\ \bibnamefont
  {Mediano}}, \bibinfo {author} {\bibfnamefont {M.}~\bibnamefont {Gastpar}},\
  and\ \bibinfo {author} {\bibfnamefont {H.~J.}\ \bibnamefont {Jensen}},\
  }\bibfield  {title} {\bibinfo {title} {Quantifying high-order
  interdependencies via multivariate extensions of the mutual information},\
  }\href {https://doi.org/10.1103/PhysRevE.100.032305} {\bibfield  {journal}
  {\bibinfo  {journal} {Phys. Rev. E}\ }\textbf {\bibinfo {volume} {100}},\
  \bibinfo {pages} {032305} (\bibinfo {year} {2019})}\BibitemShut {NoStop}%
\bibitem [{\citenamefont {Ehrlich}\ \emph {et~al.}(2023)\citenamefont
  {Ehrlich}, \citenamefont {Schneider}, \citenamefont {Priesemann},
  \citenamefont {Wibral},\ and\ \citenamefont
  {Makkeh}}]{ehrlich2023complexity}%
  \BibitemOpen
  \bibfield  {author} {\bibinfo {author} {\bibfnamefont {D.~A.}\ \bibnamefont
  {Ehrlich}}, \bibinfo {author} {\bibfnamefont {A.~C.}\ \bibnamefont
  {Schneider}}, \bibinfo {author} {\bibfnamefont {V.}~\bibnamefont
  {Priesemann}}, \bibinfo {author} {\bibfnamefont {M.}~\bibnamefont {Wibral}},\
  and\ \bibinfo {author} {\bibfnamefont {A.}~\bibnamefont {Makkeh}},\
  }\bibfield  {title} {\bibinfo {title} {A measure of the complexity of neural
  representations based on partial information decomposition},\ }\href
  {https://openreview.net/forum?id=R8TU3pfzFr} {\bibfield  {journal} {\bibinfo
  {journal} {Transactions on Machine Learning Research}\ } (\bibinfo {year}
  {2023})}\BibitemShut {NoStop}%
\bibitem [{\citenamefont {Martignon}\ \emph {et~al.}(2000)\citenamefont
  {Martignon}, \citenamefont {Deco}, \citenamefont {Laskey}, \citenamefont
  {Diamond}, \citenamefont {Freiwald},\ and\ \citenamefont
  {Vaadia}}]{martignon2000neural}%
  \BibitemOpen
  \bibfield  {author} {\bibinfo {author} {\bibfnamefont {L.}~\bibnamefont
  {Martignon}}, \bibinfo {author} {\bibfnamefont {G.}~\bibnamefont {Deco}},
  \bibinfo {author} {\bibfnamefont {K.}~\bibnamefont {Laskey}}, \bibinfo
  {author} {\bibfnamefont {M.}~\bibnamefont {Diamond}}, \bibinfo {author}
  {\bibfnamefont {W.}~\bibnamefont {Freiwald}},\ and\ \bibinfo {author}
  {\bibfnamefont {E.}~\bibnamefont {Vaadia}},\ }\bibfield  {title} {\bibinfo
  {title} {Neural coding: higher-order temporal patterns in the neurostatistics
  of cell assemblies},\ }\href@noop {} {\bibfield  {journal} {\bibinfo
  {journal} {Neural computation}\ }\textbf {\bibinfo {volume} {12}},\ \bibinfo
  {pages} {2621} (\bibinfo {year} {2000})}\BibitemShut {NoStop}%
\bibitem [{\citenamefont {Wibral}\ \emph {et~al.}(2017)\citenamefont {Wibral},
  \citenamefont {Finn}, \citenamefont {Wollstadt}, \citenamefont {Lizier},\
  and\ \citenamefont {Priesemann}}]{wibral2017quantifying}%
  \BibitemOpen
  \bibfield  {author} {\bibinfo {author} {\bibfnamefont {M.}~\bibnamefont
  {Wibral}}, \bibinfo {author} {\bibfnamefont {C.}~\bibnamefont {Finn}},
  \bibinfo {author} {\bibfnamefont {P.}~\bibnamefont {Wollstadt}}, \bibinfo
  {author} {\bibfnamefont {J.~T.}\ \bibnamefont {Lizier}},\ and\ \bibinfo
  {author} {\bibfnamefont {V.}~\bibnamefont {Priesemann}},\ }\bibfield  {title}
  {\bibinfo {title} {Quantifying information modification in developing neural
  networks via partial information decomposition},\ }\href@noop {} {\bibfield
  {journal} {\bibinfo  {journal} {Entropy}\ }\textbf {\bibinfo {volume} {19}},\
  \bibinfo {pages} {494} (\bibinfo {year} {2017})}\BibitemShut {NoStop}%
\bibitem [{\citenamefont {Varley}\ \emph {et~al.}(2023)\citenamefont {Varley},
  \citenamefont {Pope}, \citenamefont {Grazia}, \citenamefont {Joshua},\ and\
  \citenamefont {Sporns}}]{varley2023partial}%
  \BibitemOpen
  \bibfield  {author} {\bibinfo {author} {\bibfnamefont {T.~F.}\ \bibnamefont
  {Varley}}, \bibinfo {author} {\bibfnamefont {M.}~\bibnamefont {Pope}},
  \bibinfo {author} {\bibfnamefont {M.}~\bibnamefont {Grazia}}, \bibinfo
  {author} {\bibnamefont {Joshua}},\ and\ \bibinfo {author} {\bibfnamefont
  {O.}~\bibnamefont {Sporns}},\ }\bibfield  {title} {\bibinfo {title} {Partial
  entropy decomposition reveals higher-order information structures in human
  brain activity},\ }\href@noop {} {\bibfield  {journal} {\bibinfo  {journal}
  {Proceedings of the National Academy of Sciences}\ }\textbf {\bibinfo
  {volume} {120}},\ \bibinfo {pages} {e2300888120} (\bibinfo {year}
  {2023})}\BibitemShut {NoStop}%
\bibitem [{\citenamefont {Luppi}\ \emph {et~al.}(2024)\citenamefont {Luppi},
  \citenamefont {Rosas}, \citenamefont {Mediano}, \citenamefont {Menon},\ and\
  \citenamefont {Stamatakis}}]{luppi2024pidbrain}%
  \BibitemOpen
  \bibfield  {author} {\bibinfo {author} {\bibfnamefont {A.~I.}\ \bibnamefont
  {Luppi}}, \bibinfo {author} {\bibfnamefont {F.~E.}\ \bibnamefont {Rosas}},
  \bibinfo {author} {\bibfnamefont {P.~A.}\ \bibnamefont {Mediano}}, \bibinfo
  {author} {\bibfnamefont {D.~K.}\ \bibnamefont {Menon}},\ and\ \bibinfo
  {author} {\bibfnamefont {E.~A.}\ \bibnamefont {Stamatakis}},\ }\bibfield
  {title} {\bibinfo {title} {Information decomposition and the informational
  architecture of the brain},\ }\href@noop {} {\bibfield  {journal} {\bibinfo
  {journal} {Trends in Cognitive Sciences}\ } (\bibinfo {year}
  {2024})}\BibitemShut {NoStop}%
\bibitem [{\citenamefont {Pope}\ \emph {et~al.}(2024)\citenamefont {Pope},
  \citenamefont {Varley},\ and\ \citenamefont {Sporns}}]{pope2024time}%
  \BibitemOpen
  \bibfield  {author} {\bibinfo {author} {\bibfnamefont {M.}~\bibnamefont
  {Pope}}, \bibinfo {author} {\bibfnamefont {T.~F.}\ \bibnamefont {Varley}},\
  and\ \bibinfo {author} {\bibfnamefont {O.}~\bibnamefont {Sporns}},\
  }\bibfield  {title} {\bibinfo {title} {Time-varying synergy/redundancy
  dominance in the human cerebral cortex},\ }\href@noop {} {\bibfield
  {journal} {\bibinfo  {journal} {bioRxiv}\ ,\ \bibinfo {pages} {2024}}
  (\bibinfo {year} {2024})}\BibitemShut {NoStop}%
\bibitem [{\citenamefont {Miller}\ \emph {et~al.}(2014)\citenamefont {Miller},
  \citenamefont {Wang}, \citenamefont {Lizier}, \citenamefont {Prokopenko},\
  and\ \citenamefont {Rossi}}]{miller2014measuring}%
  \BibitemOpen
  \bibfield  {author} {\bibinfo {author} {\bibfnamefont {J.~M.}\ \bibnamefont
  {Miller}}, \bibinfo {author} {\bibfnamefont {X.~R.}\ \bibnamefont {Wang}},
  \bibinfo {author} {\bibfnamefont {J.~T.}\ \bibnamefont {Lizier}}, \bibinfo
  {author} {\bibfnamefont {M.}~\bibnamefont {Prokopenko}},\ and\ \bibinfo
  {author} {\bibfnamefont {L.~F.}\ \bibnamefont {Rossi}},\ }\bibfield  {title}
  {\bibinfo {title} {Measuring information dynamics in swarms},\ }in\
  \href@noop {} {\emph {\bibinfo {booktitle} {Guided self-organization:
  Inception}}}\ (\bibinfo  {publisher} {Springer},\ \bibinfo {year} {2014})\
  pp.\ \bibinfo {pages} {343--364}\BibitemShut {NoStop}%
\bibitem [{\citenamefont {Pilkiewicz}\ \emph {et~al.}(2020)\citenamefont
  {Pilkiewicz}, \citenamefont {Lemasson}, \citenamefont {Rowland},
  \citenamefont {Hein}, \citenamefont {Sun}, \citenamefont {Berdahl},
  \citenamefont {Mayo}, \citenamefont {Moehlis}, \citenamefont {Porfiri},
  \citenamefont {Fern{\'a}ndez-Juricic} \emph
  {et~al.}}]{pilkiewicz2020decoding}%
  \BibitemOpen
  \bibfield  {author} {\bibinfo {author} {\bibfnamefont {K.}~\bibnamefont
  {Pilkiewicz}}, \bibinfo {author} {\bibfnamefont {B.}~\bibnamefont
  {Lemasson}}, \bibinfo {author} {\bibfnamefont {M.}~\bibnamefont {Rowland}},
  \bibinfo {author} {\bibfnamefont {A.}~\bibnamefont {Hein}}, \bibinfo {author}
  {\bibfnamefont {J.}~\bibnamefont {Sun}}, \bibinfo {author} {\bibfnamefont
  {A.}~\bibnamefont {Berdahl}}, \bibinfo {author} {\bibfnamefont
  {M.}~\bibnamefont {Mayo}}, \bibinfo {author} {\bibfnamefont {J.}~\bibnamefont
  {Moehlis}}, \bibinfo {author} {\bibfnamefont {M.}~\bibnamefont {Porfiri}},
  \bibinfo {author} {\bibfnamefont {E.}~\bibnamefont {Fern{\'a}ndez-Juricic}},
  \emph {et~al.},\ }\bibfield  {title} {\bibinfo {title} {Decoding collective
  communications using information theory tools},\ }\href@noop {} {\bibfield
  {journal} {\bibinfo  {journal} {Journal of the Royal Society Interface}\
  }\textbf {\bibinfo {volume} {17}},\ \bibinfo {pages} {20190563} (\bibinfo
  {year} {2020})}\BibitemShut {NoStop}%
\bibitem [{\citenamefont {Twomey}\ \emph {et~al.}(2021)\citenamefont {Twomey},
  \citenamefont {Hartnett}, \citenamefont {Sosna},\ and\ \citenamefont
  {Romanczuk}}]{twomey2021redundancy}%
  \BibitemOpen
  \bibfield  {author} {\bibinfo {author} {\bibfnamefont {C.~R.}\ \bibnamefont
  {Twomey}}, \bibinfo {author} {\bibfnamefont {A.~T.}\ \bibnamefont
  {Hartnett}}, \bibinfo {author} {\bibfnamefont {M.~M.}\ \bibnamefont
  {Sosna}},\ and\ \bibinfo {author} {\bibfnamefont {P.}~\bibnamefont
  {Romanczuk}},\ }\bibfield  {title} {\bibinfo {title} {Searching for structure
  in collective systems},\ }\href@noop {} {\bibfield  {journal} {\bibinfo
  {journal} {Theory in Biosciences}\ }\textbf {\bibinfo {volume} {140}},\
  \bibinfo {pages} {361} (\bibinfo {year} {2021})}\BibitemShut {NoStop}%
\bibitem [{\citenamefont {Burns}\ \emph {et~al.}(2022)\citenamefont {Burns},
  \citenamefont {Schaerf}, \citenamefont {Lizier}, \citenamefont {Kawaguchi},
  \citenamefont {Cox}, \citenamefont {King}, \citenamefont {Krause},\ and\
  \citenamefont {Ward}}]{burns2022self}%
  \BibitemOpen
  \bibfield  {author} {\bibinfo {author} {\bibfnamefont {A.~L.}\ \bibnamefont
  {Burns}}, \bibinfo {author} {\bibfnamefont {T.~M.}\ \bibnamefont {Schaerf}},
  \bibinfo {author} {\bibfnamefont {J.}~\bibnamefont {Lizier}}, \bibinfo
  {author} {\bibfnamefont {S.}~\bibnamefont {Kawaguchi}}, \bibinfo {author}
  {\bibfnamefont {M.}~\bibnamefont {Cox}}, \bibinfo {author} {\bibfnamefont
  {R.}~\bibnamefont {King}}, \bibinfo {author} {\bibfnamefont {J.}~\bibnamefont
  {Krause}},\ and\ \bibinfo {author} {\bibfnamefont {A.~J.}\ \bibnamefont
  {Ward}},\ }\bibfield  {title} {\bibinfo {title} {Self-organization and
  information transfer in antarctic krill swarms},\ }\href@noop {} {\bibfield
  {journal} {\bibinfo  {journal} {Proceedings of the Royal Society B}\ }\textbf
  {\bibinfo {volume} {289}},\ \bibinfo {pages} {20212361} (\bibinfo {year}
  {2022})}\BibitemShut {NoStop}%
\bibitem [{\citenamefont {Chan}\ \emph {et~al.}(2017)\citenamefont {Chan},
  \citenamefont {Stumpf},\ and\ \citenamefont {Babtie}}]{chan2017gene}%
  \BibitemOpen
  \bibfield  {author} {\bibinfo {author} {\bibfnamefont {T.~E.}\ \bibnamefont
  {Chan}}, \bibinfo {author} {\bibfnamefont {M.~P.}\ \bibnamefont {Stumpf}},\
  and\ \bibinfo {author} {\bibfnamefont {A.~C.}\ \bibnamefont {Babtie}},\
  }\bibfield  {title} {\bibinfo {title} {Gene regulatory network inference from
  single-cell data using multivariate information measures},\ }\href@noop {}
  {\bibfield  {journal} {\bibinfo  {journal} {Cell systems}\ }\textbf {\bibinfo
  {volume} {5}},\ \bibinfo {pages} {251} (\bibinfo {year} {2017})}\BibitemShut
  {NoStop}%
\bibitem [{\citenamefont {Sootla}\ \emph {et~al.}(2017)\citenamefont {Sootla},
  \citenamefont {Theis},\ and\ \citenamefont {Vicente}}]{sootla2017analyzing}%
  \BibitemOpen
  \bibfield  {author} {\bibinfo {author} {\bibfnamefont {S.}~\bibnamefont
  {Sootla}}, \bibinfo {author} {\bibfnamefont {D.~O.}\ \bibnamefont {Theis}},\
  and\ \bibinfo {author} {\bibfnamefont {R.}~\bibnamefont {Vicente}},\
  }\bibfield  {title} {\bibinfo {title} {Analyzing information distribution in
  complex systems},\ }\href@noop {} {\bibfield  {journal} {\bibinfo  {journal}
  {Entropy}\ }\textbf {\bibinfo {volume} {19}},\ \bibinfo {pages} {636}
  (\bibinfo {year} {2017})}\BibitemShut {NoStop}%
\bibitem [{\citenamefont {Scagliarini}\ \emph {et~al.}(2023)\citenamefont
  {Scagliarini}, \citenamefont {Nuzzi}, \citenamefont {Antonacci},
  \citenamefont {Faes}, \citenamefont {Rosas}, \citenamefont {Marinazzo},\ and\
  \citenamefont {Stramaglia}}]{rosas2023ogradients}%
  \BibitemOpen
  \bibfield  {author} {\bibinfo {author} {\bibfnamefont {T.}~\bibnamefont
  {Scagliarini}}, \bibinfo {author} {\bibfnamefont {D.}~\bibnamefont {Nuzzi}},
  \bibinfo {author} {\bibfnamefont {Y.}~\bibnamefont {Antonacci}}, \bibinfo
  {author} {\bibfnamefont {L.}~\bibnamefont {Faes}}, \bibinfo {author}
  {\bibfnamefont {F.~E.}\ \bibnamefont {Rosas}}, \bibinfo {author}
  {\bibfnamefont {D.}~\bibnamefont {Marinazzo}},\ and\ \bibinfo {author}
  {\bibfnamefont {S.}~\bibnamefont {Stramaglia}},\ }\bibfield  {title}
  {\bibinfo {title} {Gradients of o-information: Low-order descriptors of
  high-order dependencies},\ }\href
  {https://doi.org/10.1103/PhysRevResearch.5.013025} {\bibfield  {journal}
  {\bibinfo  {journal} {Phys. Rev. Res.}\ }\textbf {\bibinfo {volume} {5}},\
  \bibinfo {pages} {013025} (\bibinfo {year} {2023})}\BibitemShut {NoStop}%
\bibitem [{\citenamefont {Watanabe}(1960)}]{watanabe1960information}%
  \BibitemOpen
  \bibfield  {author} {\bibinfo {author} {\bibfnamefont {S.}~\bibnamefont
  {Watanabe}},\ }\bibfield  {title} {\bibinfo {title} {Information theoretical
  analysis of multivariate correlation},\ }\href@noop {} {\bibfield  {journal}
  {\bibinfo  {journal} {IBM Journal of research and development}\ }\textbf
  {\bibinfo {volume} {4}},\ \bibinfo {pages} {66} (\bibinfo {year}
  {1960})}\BibitemShut {NoStop}%
\bibitem [{\citenamefont {James}\ \emph {et~al.}(2011)\citenamefont {James},
  \citenamefont {Ellison},\ and\ \citenamefont
  {Crutchfield}}]{james2011anatomy}%
  \BibitemOpen
  \bibfield  {author} {\bibinfo {author} {\bibfnamefont {R.~G.}\ \bibnamefont
  {James}}, \bibinfo {author} {\bibfnamefont {C.~J.}\ \bibnamefont {Ellison}},\
  and\ \bibinfo {author} {\bibfnamefont {J.~P.}\ \bibnamefont {Crutchfield}},\
  }\bibfield  {title} {\bibinfo {title} {Anatomy of a bit: Information in a
  time series observation},\ }\href@noop {} {\bibfield  {journal} {\bibinfo
  {journal} {Chaos: An Interdisciplinary Journal of Nonlinear Science}\
  }\textbf {\bibinfo {volume} {21}} (\bibinfo {year} {2011})}\BibitemShut
  {NoStop}%
\bibitem [{\citenamefont {Marois}\ and\ \citenamefont
  {Ivanoff}(2005)}]{marois2005capacity}%
  \BibitemOpen
  \bibfield  {author} {\bibinfo {author} {\bibfnamefont {R.}~\bibnamefont
  {Marois}}\ and\ \bibinfo {author} {\bibfnamefont {J.}~\bibnamefont
  {Ivanoff}},\ }\bibfield  {title} {\bibinfo {title} {Capacity limits of
  information processing in the brain},\ }\href@noop {} {\bibfield  {journal}
  {\bibinfo  {journal} {Trends in cognitive sciences}\ }\textbf {\bibinfo
  {volume} {9}},\ \bibinfo {pages} {296} (\bibinfo {year} {2005})}\BibitemShut
  {NoStop}%
\bibitem [{\citenamefont {Williams}\ and\ \citenamefont
  {Beer}(2010)}]{williams2010PID}%
  \BibitemOpen
  \bibfield  {author} {\bibinfo {author} {\bibfnamefont {P.~L.}\ \bibnamefont
  {Williams}}\ and\ \bibinfo {author} {\bibfnamefont {R.~D.}\ \bibnamefont
  {Beer}},\ }\bibfield  {title} {\bibinfo {title} {Nonnegative decomposition of
  multivariate information},\ }\href@noop {} {\bibfield  {journal} {\bibinfo
  {journal} {arXiv preprint arXiv:1004.2515}\ } (\bibinfo {year}
  {2010})}\BibitemShut {NoStop}%
\bibitem [{\citenamefont {Ince}(2017)}]{ince2017ped}%
  \BibitemOpen
  \bibfield  {author} {\bibinfo {author} {\bibfnamefont {R.~A.}\ \bibnamefont
  {Ince}},\ }\bibfield  {title} {\bibinfo {title} {The partial entropy
  decomposition: Decomposing multivariate entropy and mutual information via
  pointwise common surprisal},\ }\href@noop {} {\bibfield  {journal} {\bibinfo
  {journal} {arXiv preprint arXiv:1702.01591}\ } (\bibinfo {year}
  {2017})}\BibitemShut {NoStop}%
\bibitem [{\citenamefont {Timme}\ \emph {et~al.}(2014)\citenamefont {Timme},
  \citenamefont {Alford}, \citenamefont {Flecker},\ and\ \citenamefont
  {Beggs}}]{timme2014SynRedReview}%
  \BibitemOpen
  \bibfield  {author} {\bibinfo {author} {\bibfnamefont {N.}~\bibnamefont
  {Timme}}, \bibinfo {author} {\bibfnamefont {W.}~\bibnamefont {Alford}},
  \bibinfo {author} {\bibfnamefont {B.}~\bibnamefont {Flecker}},\ and\ \bibinfo
  {author} {\bibfnamefont {J.~M.}\ \bibnamefont {Beggs}},\ }\bibfield  {title}
  {\bibinfo {title} {Synergy, redundancy, and multivariate information
  measures: an experimentalist’s perspective},\ }\href@noop {} {\bibfield
  {journal} {\bibinfo  {journal} {Journal of computational neuroscience}\
  }\textbf {\bibinfo {volume} {36}},\ \bibinfo {pages} {119} (\bibinfo {year}
  {2014})}\BibitemShut {NoStop}%
\bibitem [{\citenamefont {Kolchinsky}(2024)}]{kolchinsky2024pib}%
  \BibitemOpen
  \bibfield  {author} {\bibinfo {author} {\bibfnamefont {A.}~\bibnamefont
  {Kolchinsky}},\ }\bibfield  {title} {\bibinfo {title} {Partial information
  decomposition: Redundancy as information bottleneck},\ }\href@noop {}
  {\bibfield  {journal} {\bibinfo  {journal} {Entropy}\ }\textbf {\bibinfo
  {volume} {26}},\ \bibinfo {pages} {546} (\bibinfo {year} {2024})}\BibitemShut
  {NoStop}%
\bibitem [{\citenamefont {Alemi}\ \emph {et~al.}(2017)\citenamefont {Alemi},
  \citenamefont {Fischer}, \citenamefont {Dillon},\ and\ \citenamefont
  {Murphy}}]{alemiVIB2016}%
  \BibitemOpen
  \bibfield  {author} {\bibinfo {author} {\bibfnamefont {A.~A.}\ \bibnamefont
  {Alemi}}, \bibinfo {author} {\bibfnamefont {I.}~\bibnamefont {Fischer}},
  \bibinfo {author} {\bibfnamefont {J.~V.}\ \bibnamefont {Dillon}},\ and\
  \bibinfo {author} {\bibfnamefont {K.}~\bibnamefont {Murphy}},\ }\bibfield
  {title} {\bibinfo {title} {Deep variational information bottleneck},\ }in\
  \href {https://openreview.net/forum?id=HyxQzBceg} {\emph {\bibinfo
  {booktitle} {International Conference on Learning Representations}}}\
  (\bibinfo {year} {2017})\BibitemShut {NoStop}%
\bibitem [{\citenamefont {Koch-Janusz}\ and\ \citenamefont
  {Ringel}(2018)}]{koch2018natphys}%
  \BibitemOpen
  \bibfield  {author} {\bibinfo {author} {\bibfnamefont {M.}~\bibnamefont
  {Koch-Janusz}}\ and\ \bibinfo {author} {\bibfnamefont {Z.}~\bibnamefont
  {Ringel}},\ }\bibfield  {title} {\bibinfo {title} {Mutual information, neural
  networks and the renormalization group},\ }\href@noop {} {\bibfield
  {journal} {\bibinfo  {journal} {Nature Physics}\ }\textbf {\bibinfo {volume}
  {14}},\ \bibinfo {pages} {578} (\bibinfo {year} {2018})}\BibitemShut
  {NoStop}%
\bibitem [{\citenamefont {Murphy}\ and\ \citenamefont
  {Bassett}(2023)}]{dib_iclr}%
  \BibitemOpen
  \bibfield  {author} {\bibinfo {author} {\bibfnamefont {K.~A.}\ \bibnamefont
  {Murphy}}\ and\ \bibinfo {author} {\bibfnamefont {D.~S.}\ \bibnamefont
  {Bassett}},\ }\bibfield  {title} {\bibinfo {title} {Interpretability with
  full complexity by constraining feature information},\ }in\ \href
  {https://openreview.net/forum?id=R_OL5mLhsv} {\emph {\bibinfo {booktitle}
  {International Conference on Learning Representations ({ICLR})}}}\ (\bibinfo
  {year} {2023})\BibitemShut {NoStop}%
\bibitem [{\citenamefont {Murphy}\ and\ \citenamefont
  {Bassett}(2024{\natexlab{a}})}]{dib_pnas}%
  \BibitemOpen
  \bibfield  {author} {\bibinfo {author} {\bibfnamefont {K.~A.}\ \bibnamefont
  {Murphy}}\ and\ \bibinfo {author} {\bibfnamefont {D.~S.}\ \bibnamefont
  {Bassett}},\ }\bibfield  {title} {\bibinfo {title} {Information decomposition
  in complex systems via machine learning},\ }\href@noop {} {\bibfield
  {journal} {\bibinfo  {journal} {Proceedings of the National Academy of
  Sciences}\ }\textbf {\bibinfo {volume} {121}},\ \bibinfo {pages}
  {e2312988121} (\bibinfo {year} {2024}{\natexlab{a}})}\BibitemShut {NoStop}%
\bibitem [{\citenamefont {Murphy}\ and\ \citenamefont
  {Bassett}(2024{\natexlab{b}})}]{dib_chaos}%
  \BibitemOpen
  \bibfield  {author} {\bibinfo {author} {\bibfnamefont {K.~A.}\ \bibnamefont
  {Murphy}}\ and\ \bibinfo {author} {\bibfnamefont {D.~S.}\ \bibnamefont
  {Bassett}},\ }\bibfield  {title} {\bibinfo {title} {Machine-learning
  optimized measurements of chaotic dynamical systems via the information
  bottleneck},\ }\href {https://doi.org/10.1103/PhysRevLett.132.197201}
  {\bibfield  {journal} {\bibinfo  {journal} {Phys. Rev. Lett.}\ }\textbf
  {\bibinfo {volume} {132}},\ \bibinfo {pages} {197201} (\bibinfo {year}
  {2024}{\natexlab{b}})}\BibitemShut {NoStop}%
\bibitem [{\citenamefont {Cover}\ and\ \citenamefont
  {Thomas}(1999)}]{cover1999elements}%
  \BibitemOpen
  \bibfield  {author} {\bibinfo {author} {\bibfnamefont {T.~M.}\ \bibnamefont
  {Cover}}\ and\ \bibinfo {author} {\bibfnamefont {J.~A.}\ \bibnamefont
  {Thomas}},\ }\href@noop {} {\emph {\bibinfo {title} {Elements of information
  theory}}}\ (\bibinfo  {publisher} {John Wiley \& Sons},\ \bibinfo {year}
  {1999})\BibitemShut {NoStop}%
\bibitem [{\citenamefont {Burgess}\ \emph {et~al.}(2018)\citenamefont
  {Burgess}, \citenamefont {Higgins}, \citenamefont {Pal}, \citenamefont
  {Matthey}, \citenamefont {Watters}, \citenamefont {Desjardins},\ and\
  \citenamefont {Lerchner}}]{burgess2018understanding}%
  \BibitemOpen
  \bibfield  {author} {\bibinfo {author} {\bibfnamefont {C.~P.}\ \bibnamefont
  {Burgess}}, \bibinfo {author} {\bibfnamefont {I.}~\bibnamefont {Higgins}},
  \bibinfo {author} {\bibfnamefont {A.}~\bibnamefont {Pal}}, \bibinfo {author}
  {\bibfnamefont {L.}~\bibnamefont {Matthey}}, \bibinfo {author} {\bibfnamefont
  {N.}~\bibnamefont {Watters}}, \bibinfo {author} {\bibfnamefont
  {G.}~\bibnamefont {Desjardins}},\ and\ \bibinfo {author} {\bibfnamefont
  {A.}~\bibnamefont {Lerchner}},\ }\bibfield  {title} {\bibinfo {title}
  {Understanding disentangling in $\beta$-{VAE}},\ }\href@noop {} {\bibfield
  {journal} {\bibinfo  {journal} {arXiv preprint arXiv:1804.03599}\ } (\bibinfo
  {year} {2018})}\BibitemShut {NoStop}%
\bibitem [{\citenamefont {Poole}\ \emph {et~al.}(2019)\citenamefont {Poole},
  \citenamefont {Ozair}, \citenamefont {Van Den~Oord}, \citenamefont {Alemi},\
  and\ \citenamefont {Tucker}}]{poole2019variational}%
  \BibitemOpen
  \bibfield  {author} {\bibinfo {author} {\bibfnamefont {B.}~\bibnamefont
  {Poole}}, \bibinfo {author} {\bibfnamefont {S.}~\bibnamefont {Ozair}},
  \bibinfo {author} {\bibfnamefont {A.}~\bibnamefont {Van Den~Oord}}, \bibinfo
  {author} {\bibfnamefont {A.}~\bibnamefont {Alemi}},\ and\ \bibinfo {author}
  {\bibfnamefont {G.}~\bibnamefont {Tucker}},\ }\bibfield  {title} {\bibinfo
  {title} {On variational bounds of mutual information},\ }in\ \href
  {https://proceedings.mlr.press/v97/poole19a/poole19a.pdf} {\emph {\bibinfo
  {booktitle} {International Conference on Machine Learning}}}\ (\bibinfo
  {organization} {PMLR},\ \bibinfo {year} {2019})\ pp.\ \bibinfo {pages}
  {5171--5180}\BibitemShut {NoStop}%
\bibitem [{\citenamefont {Oord}\ \emph {et~al.}(2018)\citenamefont {Oord},
  \citenamefont {Li},\ and\ \citenamefont {Vinyals}}]{oord2018InfoNCE}%
  \BibitemOpen
  \bibfield  {author} {\bibinfo {author} {\bibfnamefont {A.~v.~d.}\
  \bibnamefont {Oord}}, \bibinfo {author} {\bibfnamefont {Y.}~\bibnamefont
  {Li}},\ and\ \bibinfo {author} {\bibfnamefont {O.}~\bibnamefont {Vinyals}},\
  }\bibfield  {title} {\bibinfo {title} {Representation learning with
  contrastive predictive coding},\ }\href@noop {} {\bibfield  {journal}
  {\bibinfo  {journal} {arXiv preprint arXiv:1807.03748}\ } (\bibinfo {year}
  {2018})}\BibitemShut {NoStop}%
\bibitem [{\citenamefont {Chen}\ \emph {et~al.}(2020)\citenamefont {Chen},
  \citenamefont {Kornblith}, \citenamefont {Norouzi},\ and\ \citenamefont
  {Hinton}}]{chen2020simclr}%
  \BibitemOpen
  \bibfield  {author} {\bibinfo {author} {\bibfnamefont {T.}~\bibnamefont
  {Chen}}, \bibinfo {author} {\bibfnamefont {S.}~\bibnamefont {Kornblith}},
  \bibinfo {author} {\bibfnamefont {M.}~\bibnamefont {Norouzi}},\ and\ \bibinfo
  {author} {\bibfnamefont {G.}~\bibnamefont {Hinton}},\ }\bibfield  {title}
  {\bibinfo {title} {A simple framework for contrastive learning of visual
  representations},\ }in\ \href@noop {} {\emph {\bibinfo {booktitle}
  {International conference on machine learning}}}\ (\bibinfo {organization}
  {PMLR},\ \bibinfo {year} {2020})\ pp.\ \bibinfo {pages}
  {1597--1607}\BibitemShut {NoStop}%
\bibitem [{\citenamefont {Gomes}\ \emph {et~al.}(2000)\citenamefont {Gomes},
  \citenamefont {Selman}, \citenamefont {Crato},\ and\ \citenamefont
  {Kautz}}]{gomes2000heavy}%
  \BibitemOpen
  \bibfield  {author} {\bibinfo {author} {\bibfnamefont {C.~P.}\ \bibnamefont
  {Gomes}}, \bibinfo {author} {\bibfnamefont {B.}~\bibnamefont {Selman}},
  \bibinfo {author} {\bibfnamefont {N.}~\bibnamefont {Crato}},\ and\ \bibinfo
  {author} {\bibfnamefont {H.}~\bibnamefont {Kautz}},\ }\bibfield  {title}
  {\bibinfo {title} {Heavy-tailed phenomena in satisfiability and constraint
  satisfaction problems},\ }\href@noop {} {\bibfield  {journal} {\bibinfo
  {journal} {Journal of automated reasoning}\ }\textbf {\bibinfo {volume}
  {24}},\ \bibinfo {pages} {67} (\bibinfo {year} {2000})}\BibitemShut {NoStop}%
\bibitem [{\citenamefont {Ercsey-Ravasz}\ and\ \citenamefont
  {Toroczkai}(2012)}]{ercsey2012chaos}%
  \BibitemOpen
  \bibfield  {author} {\bibinfo {author} {\bibfnamefont {M.}~\bibnamefont
  {Ercsey-Ravasz}}\ and\ \bibinfo {author} {\bibfnamefont {Z.}~\bibnamefont
  {Toroczkai}},\ }\bibfield  {title} {\bibinfo {title} {The chaos within
  sudoku},\ }\href@noop {} {\bibfield  {journal} {\bibinfo  {journal}
  {Scientific reports}\ }\textbf {\bibinfo {volume} {2}},\ \bibinfo {pages}
  {725} (\bibinfo {year} {2012})}\BibitemShut {NoStop}%
\bibitem [{\citenamefont {Varga}\ \emph {et~al.}(2016)\citenamefont {Varga},
  \citenamefont {Sumi}, \citenamefont {Toroczkai},\ and\ \citenamefont
  {Ercsey-Ravasz}}]{varga2016order}%
  \BibitemOpen
  \bibfield  {author} {\bibinfo {author} {\bibfnamefont {M.}~\bibnamefont
  {Varga}}, \bibinfo {author} {\bibfnamefont {R.}~\bibnamefont {Sumi}},
  \bibinfo {author} {\bibfnamefont {Z.}~\bibnamefont {Toroczkai}},\ and\
  \bibinfo {author} {\bibfnamefont {M.}~\bibnamefont {Ercsey-Ravasz}},\
  }\bibfield  {title} {\bibinfo {title} {Order-to-chaos transition in the
  hardness of random boolean satisfiability problems},\ }\href@noop {}
  {\bibfield  {journal} {\bibinfo  {journal} {Physical Review E}\ }\textbf
  {\bibinfo {volume} {93}},\ \bibinfo {pages} {052211} (\bibinfo {year}
  {2016})}\BibitemShut {NoStop}%
\bibitem [{\citenamefont {Kailath}(1967)}]{kailath1967Bdistance}%
  \BibitemOpen
  \bibfield  {author} {\bibinfo {author} {\bibfnamefont {T.}~\bibnamefont
  {Kailath}},\ }\bibfield  {title} {\bibinfo {title} {The divergence and
  bhattacharyya distance measures in signal selection},\ }\href@noop {}
  {\bibfield  {journal} {\bibinfo  {journal} {IEEE transactions on
  communication technology}\ }\textbf {\bibinfo {volume} {15}},\ \bibinfo
  {pages} {52} (\bibinfo {year} {1967})}\BibitemShut {NoStop}%
\bibitem [{\citenamefont {Murphy}\ \emph {et~al.}(2025)\citenamefont {Murphy},
  \citenamefont {Dillavou},\ and\ \citenamefont {Bassett}}]{nmi2024}%
  \BibitemOpen
  \bibfield  {author} {\bibinfo {author} {\bibfnamefont {K.~A.}\ \bibnamefont
  {Murphy}}, \bibinfo {author} {\bibfnamefont {S.}~\bibnamefont {Dillavou}},\
  and\ \bibinfo {author} {\bibfnamefont {D.~S.}\ \bibnamefont {Bassett}},\
  }\bibfield  {title} {\bibinfo {title} {Comparing the information content of
  probabilistic representation spaces},\ }\href
  {https://openreview.net/forum?id=adhsMqURI1} {\bibfield  {journal} {\bibinfo
  {journal} {Transactions on Machine Learning Research}\ } (\bibinfo {year}
  {2025})}\BibitemShut {NoStop}%
\bibitem [{\citenamefont {Shannon}(1948)}]{shannon1948mathematical}%
  \BibitemOpen
  \bibfield  {author} {\bibinfo {author} {\bibfnamefont {C.~E.}\ \bibnamefont
  {Shannon}},\ }\bibfield  {title} {\bibinfo {title} {A mathematical theory of
  communication},\ }\href@noop {} {\bibfield  {journal} {\bibinfo  {journal}
  {The Bell system technical journal}\ }\textbf {\bibinfo {volume} {27}},\
  \bibinfo {pages} {379} (\bibinfo {year} {1948})}\BibitemShut {NoStop}%
\bibitem [{\citenamefont {Stephens}\ and\ \citenamefont
  {Bialek}(2010)}]{stephens2010letters}%
  \BibitemOpen
  \bibfield  {author} {\bibinfo {author} {\bibfnamefont {G.~J.}\ \bibnamefont
  {Stephens}}\ and\ \bibinfo {author} {\bibfnamefont {W.}~\bibnamefont
  {Bialek}},\ }\bibfield  {title} {\bibinfo {title} {Statistical mechanics of
  letters in words},\ }\href@noop {} {\bibfield  {journal} {\bibinfo  {journal}
  {Physical Review E—Statistical, Nonlinear, and Soft Matter Physics}\
  }\textbf {\bibinfo {volume} {81}},\ \bibinfo {pages} {066119} (\bibinfo
  {year} {2010})}\BibitemShut {NoStop}%
\bibitem [{\citenamefont {Scagliarini}\ \emph {et~al.}(2022)\citenamefont
  {Scagliarini}, \citenamefont {Marinazzo}, \citenamefont {Guo}, \citenamefont
  {Stramaglia},\ and\ \citenamefont {Rosas}}]{scagliarini2022local}%
  \BibitemOpen
  \bibfield  {author} {\bibinfo {author} {\bibfnamefont {T.}~\bibnamefont
  {Scagliarini}}, \bibinfo {author} {\bibfnamefont {D.}~\bibnamefont
  {Marinazzo}}, \bibinfo {author} {\bibfnamefont {Y.}~\bibnamefont {Guo}},
  \bibinfo {author} {\bibfnamefont {S.}~\bibnamefont {Stramaglia}},\ and\
  \bibinfo {author} {\bibfnamefont {F.~E.}\ \bibnamefont {Rosas}},\ }\bibfield
  {title} {\bibinfo {title} {Quantifying high-order interdependencies on
  individual patterns via the local o-information: Theory and applications to
  music analysis},\ }\href@noop {} {\bibfield  {journal} {\bibinfo  {journal}
  {Physical Review Research}\ }\textbf {\bibinfo {volume} {4}},\ \bibinfo
  {pages} {013184} (\bibinfo {year} {2022})}\BibitemShut {NoStop}%
\bibitem [{\citenamefont {Nirenberg}\ \emph {et~al.}(2001)\citenamefont
  {Nirenberg}, \citenamefont {Carcieri}, \citenamefont {Jacobs},\ and\
  \citenamefont {Latham}}]{nirenberg2001retinal}%
  \BibitemOpen
  \bibfield  {author} {\bibinfo {author} {\bibfnamefont {S.}~\bibnamefont
  {Nirenberg}}, \bibinfo {author} {\bibfnamefont {S.~M.}\ \bibnamefont
  {Carcieri}}, \bibinfo {author} {\bibfnamefont {A.~L.}\ \bibnamefont
  {Jacobs}},\ and\ \bibinfo {author} {\bibfnamefont {P.~E.}\ \bibnamefont
  {Latham}},\ }\bibfield  {title} {\bibinfo {title} {Retinal ganglion cells act
  largely as independent encoders},\ }\href@noop {} {\bibfield  {journal}
  {\bibinfo  {journal} {Nature}\ }\textbf {\bibinfo {volume} {411}},\ \bibinfo
  {pages} {698} (\bibinfo {year} {2001})}\BibitemShut {NoStop}%
\bibitem [{\citenamefont {Maliniak}\ \emph {et~al.}(2013)\citenamefont
  {Maliniak}, \citenamefont {Powers},\ and\ \citenamefont
  {Walter}}]{maliniak2013gender}%
  \BibitemOpen
  \bibfield  {author} {\bibinfo {author} {\bibfnamefont {D.}~\bibnamefont
  {Maliniak}}, \bibinfo {author} {\bibfnamefont {R.}~\bibnamefont {Powers}},\
  and\ \bibinfo {author} {\bibfnamefont {B.~F.}\ \bibnamefont {Walter}},\
  }\bibfield  {title} {\bibinfo {title} {The gender citation gap in
  international relations},\ }\href@noop {} {\bibfield  {journal} {\bibinfo
  {journal} {International Organization}\ }\textbf {\bibinfo {volume} {67}},\
  \bibinfo {pages} {889} (\bibinfo {year} {2013})}\BibitemShut {NoStop}%
\bibitem [{\citenamefont {Caplar}\ \emph {et~al.}(2017)\citenamefont {Caplar},
  \citenamefont {Tacchella},\ and\ \citenamefont
  {Birrer}}]{caplar2017quantitative}%
  \BibitemOpen
  \bibfield  {author} {\bibinfo {author} {\bibfnamefont {N.}~\bibnamefont
  {Caplar}}, \bibinfo {author} {\bibfnamefont {S.}~\bibnamefont {Tacchella}},\
  and\ \bibinfo {author} {\bibfnamefont {S.}~\bibnamefont {Birrer}},\
  }\bibfield  {title} {\bibinfo {title} {Quantitative evaluation of gender bias
  in astronomical publications from citation counts},\ }\href@noop {}
  {\bibfield  {journal} {\bibinfo  {journal} {Nature Astronomy}\ }\textbf
  {\bibinfo {volume} {1}},\ \bibinfo {pages} {1} (\bibinfo {year}
  {2017})}\BibitemShut {NoStop}%
\bibitem [{\citenamefont {Chakravartty}\ \emph {et~al.}(2018)\citenamefont
  {Chakravartty}, \citenamefont {Kuo}, \citenamefont {Grubbs},\ and\
  \citenamefont {McIlwain}}]{chakravartty2018communicationsowhite}%
  \BibitemOpen
  \bibfield  {author} {\bibinfo {author} {\bibfnamefont {P.}~\bibnamefont
  {Chakravartty}}, \bibinfo {author} {\bibfnamefont {R.}~\bibnamefont {Kuo}},
  \bibinfo {author} {\bibfnamefont {V.}~\bibnamefont {Grubbs}},\ and\ \bibinfo
  {author} {\bibfnamefont {C.}~\bibnamefont {McIlwain}},\ }\bibfield  {title}
  {\bibinfo {title} {{\#CommunicationSoWhite}},\ }\href@noop {} {\bibfield
  {journal} {\bibinfo  {journal} {Journal of Communication}\ }\textbf {\bibinfo
  {volume} {68}},\ \bibinfo {pages} {254} (\bibinfo {year} {2018})}\BibitemShut
  {NoStop}%
\bibitem [{\citenamefont {Dion}\ \emph {et~al.}(2018)\citenamefont {Dion},
  \citenamefont {Sumner},\ and\ \citenamefont {Mitchell}}]{dion2018gendered}%
  \BibitemOpen
  \bibfield  {author} {\bibinfo {author} {\bibfnamefont {M.~L.}\ \bibnamefont
  {Dion}}, \bibinfo {author} {\bibfnamefont {J.~L.}\ \bibnamefont {Sumner}},\
  and\ \bibinfo {author} {\bibfnamefont {S.~M.}\ \bibnamefont {Mitchell}},\
  }\bibfield  {title} {\bibinfo {title} {Gendered citation patterns across
  political science and social science methodology fields},\ }\href@noop {}
  {\bibfield  {journal} {\bibinfo  {journal} {Political Analysis}\ }\textbf
  {\bibinfo {volume} {26}},\ \bibinfo {pages} {312} (\bibinfo {year}
  {2018})}\BibitemShut {NoStop}%
\bibitem [{\citenamefont {Dworkin}\ \emph
  {et~al.}(2020{\natexlab{a}})\citenamefont {Dworkin}, \citenamefont {Linn},
  \citenamefont {Teich}, \citenamefont {Zurn}, \citenamefont {Shinohara},\ and\
  \citenamefont {Bassett}}]{dworkin2020extent}%
  \BibitemOpen
  \bibfield  {author} {\bibinfo {author} {\bibfnamefont {J.~D.}\ \bibnamefont
  {Dworkin}}, \bibinfo {author} {\bibfnamefont {K.~A.}\ \bibnamefont {Linn}},
  \bibinfo {author} {\bibfnamefont {E.~G.}\ \bibnamefont {Teich}}, \bibinfo
  {author} {\bibfnamefont {P.}~\bibnamefont {Zurn}}, \bibinfo {author}
  {\bibfnamefont {R.~T.}\ \bibnamefont {Shinohara}},\ and\ \bibinfo {author}
  {\bibfnamefont {D.~S.}\ \bibnamefont {Bassett}},\ }\bibfield  {title}
  {\bibinfo {title} {The extent and drivers of gender imbalance in neuroscience
  reference lists},\ }\href@noop {} {\bibfield  {journal} {\bibinfo  {journal}
  {Nature Neuroscience}\ }\textbf {\bibinfo {volume} {23}},\ \bibinfo {pages}
  {918} (\bibinfo {year} {2020}{\natexlab{a}})}\BibitemShut {NoStop}%
\bibitem [{\citenamefont {Teich}\ \emph {et~al.}(2022)\citenamefont {Teich},
  \citenamefont {Kim}, \citenamefont {Lynn}, \citenamefont {Simon},
  \citenamefont {Klishin}, \citenamefont {Szymula}, \citenamefont {Srivastava},
  \citenamefont {Bassett}, \citenamefont {Zurn}, \citenamefont {Dworkin} \emph
  {et~al.}}]{teich2022citation}%
  \BibitemOpen
  \bibfield  {author} {\bibinfo {author} {\bibfnamefont {E.~G.}\ \bibnamefont
  {Teich}}, \bibinfo {author} {\bibfnamefont {J.~Z.}\ \bibnamefont {Kim}},
  \bibinfo {author} {\bibfnamefont {C.~W.}\ \bibnamefont {Lynn}}, \bibinfo
  {author} {\bibfnamefont {S.~C.}\ \bibnamefont {Simon}}, \bibinfo {author}
  {\bibfnamefont {A.~A.}\ \bibnamefont {Klishin}}, \bibinfo {author}
  {\bibfnamefont {K.~P.}\ \bibnamefont {Szymula}}, \bibinfo {author}
  {\bibfnamefont {P.}~\bibnamefont {Srivastava}}, \bibinfo {author}
  {\bibfnamefont {L.~C.}\ \bibnamefont {Bassett}}, \bibinfo {author}
  {\bibfnamefont {P.}~\bibnamefont {Zurn}}, \bibinfo {author} {\bibfnamefont
  {J.~D.}\ \bibnamefont {Dworkin}}, \emph {et~al.},\ }\bibfield  {title}
  {\bibinfo {title} {Citation inequity and gendered citation practices in
  contemporary physics},\ }\href@noop {} {\bibfield  {journal} {\bibinfo
  {journal} {Nature Physics}\ }\textbf {\bibinfo {volume} {18}},\ \bibinfo
  {pages} {1161} (\bibinfo {year} {2022})}\BibitemShut {NoStop}%
\bibitem [{\citenamefont {Zurn}\ \emph {et~al.}(2020)\citenamefont {Zurn},
  \citenamefont {Bassett},\ and\ \citenamefont {Rust}}]{zurn2020citation}%
  \BibitemOpen
  \bibfield  {author} {\bibinfo {author} {\bibfnamefont {P.}~\bibnamefont
  {Zurn}}, \bibinfo {author} {\bibfnamefont {D.~S.}\ \bibnamefont {Bassett}},\
  and\ \bibinfo {author} {\bibfnamefont {N.~C.}\ \bibnamefont {Rust}},\
  }\bibfield  {title} {\bibinfo {title} {The citation diversity statement: a
  practice of transparency, a way of life},\ }\href@noop {} {\bibfield
  {journal} {\bibinfo  {journal} {Trends in Cognitive Sciences}\ }\textbf
  {\bibinfo {volume} {24}},\ \bibinfo {pages} {669} (\bibinfo {year}
  {2020})}\BibitemShut {NoStop}%
\bibitem [{\citenamefont {Dworkin}\ \emph
  {et~al.}(2020{\natexlab{b}})\citenamefont {Dworkin}, \citenamefont {Zurn},\
  and\ \citenamefont {Bassett}}]{dworkin2020citing}%
  \BibitemOpen
  \bibfield  {author} {\bibinfo {author} {\bibfnamefont {J.}~\bibnamefont
  {Dworkin}}, \bibinfo {author} {\bibfnamefont {P.}~\bibnamefont {Zurn}},\ and\
  \bibinfo {author} {\bibfnamefont {D.~S.}\ \bibnamefont {Bassett}},\
  }\bibfield  {title} {\bibinfo {title} {{(In)}citing action to realize an
  equitable future},\ }\href@noop {} {\bibfield  {journal} {\bibinfo  {journal}
  {Neuron}\ }\textbf {\bibinfo {volume} {106}},\ \bibinfo {pages} {890}
  (\bibinfo {year} {2020}{\natexlab{b}})}\BibitemShut {NoStop}%
\bibitem [{\citenamefont {Zhou}\ \emph {et~al.}(2020)\citenamefont {Zhou},
  \citenamefont {Cornblath}, \citenamefont {Stiso}, \citenamefont {Teich},
  \citenamefont {Dworkin}, \citenamefont {Blevins},\ and\ \citenamefont
  {Bassett}}]{zhou2020gender}%
  \BibitemOpen
  \bibfield  {author} {\bibinfo {author} {\bibfnamefont {D.}~\bibnamefont
  {Zhou}}, \bibinfo {author} {\bibfnamefont {E.~J.}\ \bibnamefont {Cornblath}},
  \bibinfo {author} {\bibfnamefont {J.}~\bibnamefont {Stiso}}, \bibinfo
  {author} {\bibfnamefont {E.~G.}\ \bibnamefont {Teich}}, \bibinfo {author}
  {\bibfnamefont {J.~D.}\ \bibnamefont {Dworkin}}, \bibinfo {author}
  {\bibfnamefont {A.~S.}\ \bibnamefont {Blevins}},\ and\ \bibinfo {author}
  {\bibfnamefont {D.~S.}\ \bibnamefont {Bassett}},\ }\bibfield  {title}
  {\bibinfo {title} {Gender diversity statement and code notebook v1. 0},\
  }\href@noop {} {\bibfield  {journal} {\bibinfo  {journal} {Zenodo}\ }
  (\bibinfo {year} {2020})}\BibitemShut {NoStop}%
\bibitem [{\citenamefont {Budrikis}(2020)}]{budrikis2020growing}%
  \BibitemOpen
  \bibfield  {author} {\bibinfo {author} {\bibfnamefont {Z.}~\bibnamefont
  {Budrikis}},\ }\bibfield  {title} {\bibinfo {title} {Growing citation gender
  gap},\ }\href@noop {} {\bibfield  {journal} {\bibinfo  {journal} {Nature
  Reviews Physics}\ }\textbf {\bibinfo {volume} {2}},\ \bibinfo {pages} {346}
  (\bibinfo {year} {2020})}\BibitemShut {NoStop}%
\end{thebibliography}
\end{document}